\documentclass[a4paper,fleqn,usenatbib]{mnras}

\usepackage[T1]{fontenc}
\usepackage{ae,aecompl}

\usepackage{multirow}
\usepackage{enumerate}
\usepackage{tablefootnote}
\usepackage{pdflscape}

\usepackage{graphicx}	
\usepackage{amsmath}	
\usepackage{amssymb}	

\title[Environmental dependence of age and metallicity gradients]{SDSS-IV MaNGA: environmental dependence of stellar age and metallicity gradients in nearby galaxies}

\author[Z. Zheng et al.]{
Zheng Zheng$^{1}$\thanks{E-mail: zz@bao.ac.cn}, 
Huiyuan Wang$^{2}$, 
Junqiang Ge$^{1}$,
Shude Mao$^{1,3,4}$,  
Cheng Li$^{3,5}$, 
\newauthor Ran Li$^{1}$,
Houjun Mo$^{3,6}$,
Daniel Goddard$^{7}$,
Kevin Bundy$^{8}$,
Hongyu Li$^{1}$,
Preethi Nair$^{9}$,
\newauthor Lihwai Lin$^{10}$,
R. J. Long$^{1,4}$,
Rog$\rm \acute{e}$rio Riffel$^{11,12}$,
Daniel Thomas$^{7}$,
Karen Masters$^{7}$,
\newauthor Dmitry Bizyaev$^{13,14}$,
Joel R. Brownstein$^{15}$,
Kai Zhang$^{16}$,
David R. Law$^{17}$,
Niv Drory$^{18}$,
\newauthor Alexandre Roman Lopes$^{19}$,
Olena Malanushenko$^{13}$
\\
$^{1}${National Astronomical Observatories, Chinese Academy of Sciences, A20 Datun Road, Chaoyang District, Beijing 100012, China} \\
$^{2}${Key Laboratory for Research in Galaxies and Cosmology, Department of Astronomy, University of Science and Technology of China,} \\ 
{    Hefei, Anhui 230026, China} \\
$^{3}${Tsinghua Center for Astrophysics and Department of Physics, Tsinghua University, Beijing 100084, China} \\
$^{4}${Jodrell Bank Centre for Astrophysics, School of Physics and Astronomy, The University of Manchester, Oxford Road, Manchester M13 9PL, UK} \\
$^{5}${Shanghai Astronomical Observatory, Chinese Academy of Sciences, Shanghai 200030, China} \\
$^{6}${Department of Astronomy, University of  Massachusetts Amherst, MA 01003, USA} \\
$^{7}${Institute of Cosmology and Gravitation, University of Portsmouth, Dennis Sciama Building, Burnaby Road, Portsmouth, PO1 3FX, UK} \\
$^{8}${Kavli IPMU (WPI), UTIAS, The University of Tokyo, Kashiwa, Chiba 277-8583, Japan} \\
$^{9}${The University of Alabama, Tuscaloosa, AL 35487, USA} \\
$^{10}${Institute of Astronomy \& Astrophysics, Academia Sinica, Taipei 10617, Taiwan} \\
$^{11}${Departamento de Astronomia, Instituto de F{\'i}sica, Universidade Federal do Rio Grande do Sul, CP 15051, 91501-970, Porto Alegre, RS, Brazil} \\
$^{12}${Laborat{\'o}rio Interinstitucional de e- Astronomia, Rua General Jos{\'e} Cristino, 77 Vasco da Gama, Rio de Janeiro, Brasil, 20921-400} \\
$^{13}${Apache Point Observatory and New Mexico State University, P.O. Box 59, Sunspot, NM, 88349-0059, USA} \\
$^{14}${Sternberg Astronomical Institute, Moscow State University, Moscow} \\
$^{15}${The University of Utah, Salt Lake City, UT 84112, USA} \\
$^{16}${University of Kentucky, Lexington, KY 40506, USA} \\
$^{17}${Space Telescope Science Institute, 3700 San Martin Drive, Baltimore, MD 21218, USA} \\
$^{18}${McDonald Observatory, The University of Texas at Austin, 1 University Station, Austin, TX 78712, USA} \\
$^{19}${Departamento de Fõsica, Facultad de Ciencias, Universidad de La Serena, Cisternas 1200, La Serena, Chile} }

\date{Accepted XXX. Received YYY; in original form ZZZ}

\pubyear{2016}

\begin{document}
\label{firstpage}
\pagerange{\pageref{firstpage}--\pageref{lastpage}}
\maketitle

\begin{abstract}

We present a study on the stellar age and metallicity distributions
for 1105 galaxies using the STARLIGHT software on MaNGA integral field
spectra. We derive  age and metallicity gradients by fitting straight
lines to the radial profiles, and explore their correlations with
total stellar mass $M_*$, $NUV-r$ colour and environments, as
identified by both the large scale structure (LSS) type and  the local
density. We find that the mean age and metallicity gradients are close
to zero but slightly negative, which is consistent with the inside-out
formation scenario.  Within our sample, we find that both the age and
metallicity gradients show weak or no correlation with either the LSS
type or local density environment. In addition, we also study the
environmental dependence of age and metallicity values at the
effective radii. The age and metallicity values are highly correlated
with $M_*$ and $NUV-r$ and are also dependent on LSS type as well as
local density. Low-mass galaxies tend to be younger and have lower
metallicity in low-density environments while high-mass galaxies are
less affected by environment.

\end{abstract}




\section{Introduction}
\label{introduction}
Galaxies are complex systems and their structures are expected to
  be affected by many different processes.  Observations have shown
  that the stellar age and metallicity gradients of galaxies are
  correlated with  the overall properties of galaxies, such as total
  stellar mass, broad band colour, and stellar velocity dispersion
  \citep[e.g.][]{mehlert03, spolaor09,koleva09,  rawle10, kuntschner10, lb12,
    gd15}. Theoretically, internal processes such as  supernova
  feedback \citep[e.g.][]{kg03} and migration of stars
  \citep[e.g.][]{sb02, roskar08} have been proposed to interpret  the
  observed gradients. In the hierarchical galaxy formation scenario,
galaxies are formed in dark matter haloes and could experience many
mergers during their formation history. Galaxy properties could
therefore be  affected also by  galaxy environments  and merger
history. Previous studies have also showed that galaxy properties
related to mass and star  formation history, e.g. total stellar mass
\citep{kauffmann04, li06}, colour \citep{blanton05, li06}, D4000
\cite[the break at 4000\AA,][]{kauffmann04, li06}  and morphology
\citep{dressler80}, are strongly dependent on environment. Structure
related parameters,  such as concentration, surface brightness and
S$\rm \acute{e}$rsic indices, at given total stellar mass and colour,
are however, nearly independent of environment
\citep{kauffmann04, blanton05, li06, bm09}.  

There have been a lot of studies about relationships between galaxy properties and local density environment using various kinds of environment indicators, such as group environment \citep{yang07}, nearest neighbour distance \citep{park07}, and number counts of neighbouring galaxies \citep{kauffmann04,bm09}.  Most previous local density environment indicators are derived using galaxies, which may be a biased tracer of the underlying mass distribution. There have  also been  studies about the dependence of galaxy overall properties on large scale structure (LSS) environments \cite[e.g.][]{ll08}.

It is natural to ask whether the distribution of star formation history related parameters, such as stellar age and metallicity, depends on galaxy environments.
On  one hand, there are many numerical simulations studying the effects of mergers on stellar population distributions \cite[e.g.][]{dm09,pipino10}.
Generally speaking, monolithic collapse models \citep{pipino10} usually produce relatively large metallicity gradients, whilst mergers  tend to make the profiles  flatter \citep{dm09}. However, secular evolution mechanisms such as stellar migrations \cite[e.g.][]{roskar08, sb09, minchev12} could also have large effects on stellar population distributions. 
On the other hand, there have been more and more resolved spectroscopic studies of stellar population distributions in  galaxies, 
\cite[e.g.][]{yoachim10,sb14,sb14b,gd15,morelli15},
but most of them have sample sizes of fewer than a few tens and none of them have explored the environmental dependence. Some photometric studies \cite[e.g.][]{zheng15, tortora10, tortora12} and some spectroscopic studies using special techniques \citep{roig15}  have large samples but their results are  affected either by the age-metallicity degeneracy or by poor spatial resolution.

Thanks to the Sloan Digital Sky Survey (SDSS) Mapping Nearby Galaxies at APO \cite[MaNGA,][]{bundy15} project, we are able to  measure integrated field unit (IFU) spectra of 10,000 galaxies and answer the question posed in the last paragraph. 
Here in this paper, we use a method \citep{wang09,wang12} based on reconstructed density field, instead of neighbouring galaxies, to derive the LSS environment information, and explore LSS environmental dependence of the stellar age and metallicity gradients. To do this, we use IFU spectra of more than 1000 galaxies from the MaNGA MPL-4 data release. 
In another two companion MaNGA papers by Goddard et al. (2016ab, submitted),  the dependence on galaxy properties and  local density environment  based  on neighbour counts is considered. 

The outline of the paper is as follows: we describe the data and galaxy sample in Section \ref{datasample};
we briefly introduce the methods for determining the LSS environment and stellar population parameters in Section \ref{methods};  we then present our results in Section \ref{results}, and discuss the results in Section \ref{discussion}; and finally we summarize our conclusions in Section \ref{conclusions}.

\section{Data and sample}
\label{datasample}

\begin{figure}
\begin{center}
\includegraphics[scale=0.4]{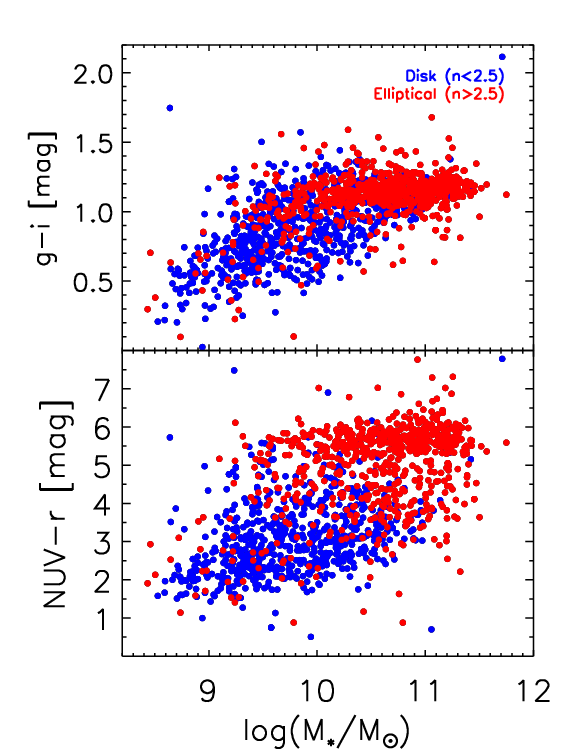}  
\caption{Mass-colour distribution of our sample. Red dots are for galaxies with S$\rm \acute{e}$rsic indices $n \ge 2.5$, which are more elliptical-like, so we call them  ``elliptical" galaxies for short. Blue dots are for galaxies with $n < 2.5$, which are more disk-like, so we call them ``disk" galaxies for short.  The NUV, SDSS $g$, $r$ and  $i$ magnitudes, and total stellar mass $M_*$ are from the NSA catalogue.
}
\label{masscolourplot}
\end{center}
\end{figure}

\begin{figure*} 
\begin{center}

\includegraphics[scale=0.5]{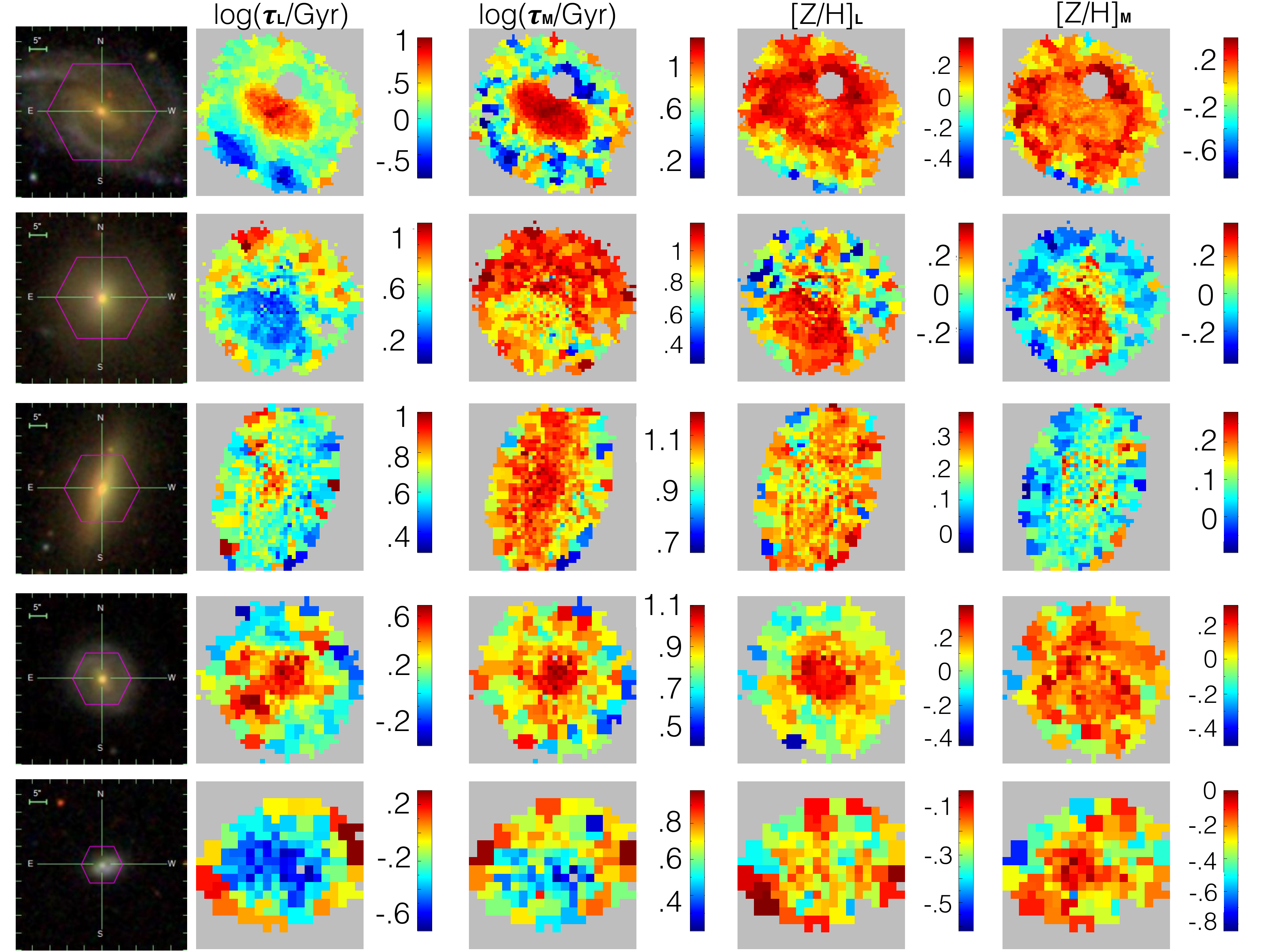}  

\caption{SDSS 3-colour images and 2D maps of STARLIGHT derived parameters of 5 example galaxies. From left to right: SDSS 3-colour image, luminosity-weighted age ($\log(\tau_L/Gyr)$) map, mass-weighted age ($\log(\tau_M/Gyr)$) map, luminosity-weighted metallicity ($\rm [Z/H]_L$) map, and mass-weighted metallicity ($\rm [Z/H]_M$)  map. The purple hexagons on the 3-colour images show the fields of view of MaNGA observations.  From top to bottom: A face-on spiral galaxy (MaNGA plate-IFU identifier: 8135-12701), an elliptical galaxy (8549-9101), an edge-on early-type galaxy (8612-6101), a face-on spiral galaxy with a bar (8549-3703), and an irregular galaxy (8549-1901). These five galaxies are observed using 127, 91, 61, 37 and 19 fibres (from top to bottom) respectively. Note that low signal-to-noise spaxels are removed from the 2D maps.
}
\label{2dmaps}
\end{center}
\end{figure*}

MaNGA \citep{bundy15,yan16} is an IFU survey targeting about 10,000 nearby galaxies selected from the SDSS.   The wavelength coverage is between 3600 \AA \, and 10300 \AA \, with a spectral resolution  R$\sim$2000 \citep{drory15}. The sizes of the IFUs vary for different galaxies from $12''$ for a 19-fibre IFU to $32''$ for a 127-fibre IFU \citep{drory15,law15}.  
The MaNGA internal data release MPL-4 sample \citep[equivalent to SDSS DR13 public release,][]{sdssdr13} has a redshift range $0.01<z<0.15$ with a peak around $z=0.03$.
Figure \ref{masscolourplot} shows the distributions of  $NUV-r$ colour and $g-r$ colour versus the total stellar mass $M_*$ of our sample galaxies.
The $M_*$, $NUV-r$ colour, $g-r$ colour, redshift and $n$ are obtained from the NASA Sloan 
Atlas\footnote{http://nsatlas.org} catalogue. The total stellar masses are derived from 
the fit to the SDSS five-band photometry with $K$-corrections
 \citep{br07,blanton11}  using the \citet{bc03} model and the \citet{chabrier03} initial mass function (IMF). The S$\rm \acute{e}$rsic indices are derived using 2D  S$\rm \acute{e}$rsic fits to the r-band SDSS images.
A few example galaxies observed with different IFU bundle sizes are shown in Fig. \ref{2dmaps}. Detailed descriptions about the derivation of stellar population parameters shown in this figure can be found in Section \ref{spsmodel}. 

The MaNGA sample is composed of primary sample, colour-enhanced sample and secondary sample.
The main MaNGA sample galaxies are selected to lie within a redshift range, $z_{min} < z < z_{max}$,
that depends on the absolute $i$-band magnitude ($M_i$) in the case of the primary and secondary 
samples, and on $M_i$ and the $NUV-r$ colour in the case of the colour-enhanced 
sample. The values of $z_{min}$ and $z_{max}$ are chosen so that both the number density of galaxies 
and the angular size distribution, matched to the IFU sizes, are roughly independent of $M_i$ 
(or $M_i$ and $NUV-r$ for the colour-enhanced sample). 
This results in lower and narrower redshift ranges for less luminous galaxies 
and higher and wider redshift ranges for more massive galaxies.
At a given $M_i$ (or $M_i$ and $NUV-r$ colour for the colour-enhanced sample) the sample is effectively 
volume limited, so that all galaxies with $z_{min} < z < z_{max}$ are targeted independently of their other 
properties.\footnote{This volume varies with $M_i$, and therefore a correction for this varying  
volume of selection is needed when calculating statistical values from the sample. We correct the galaxies back to a volume limited sample by applying a weight (W) to each galaxy such that $W = V_r/V_s$, where $V_r$ is an arbitrary reference volume.}
Two thirds of the MaNGA sample (the primary sample + the colour-enhanced sample)  are  covered by IFU observations up to 1.5 effective radii ($R_e$) and the other  third of the sample (the secondary sample)  are covered up to 2.5 $R_e$. \citep[][Wake et al. in prep.]{yan16}. Each target is observed using three dithers \citep{law15} and the observed data are reduced by the MaNGA data reduction pipeline \citep[DRP;][]{law16, yan15}. The current version of the DRP is 1\_5\_1 and the final products of the DRP are datacubes with a pixel size of $0.5''$ as well as row stacked spectra.

Our sample galaxies are taken from the MaNGA internal data release MPL-4 comprising 1351 unique galaxies.  Of these, 1144 galaxies have large scale environment information (see Section \ref{methods} for details).  Furthermore, we  remove  39 galaxies that either are too small for gradients to be derived or have multiple galaxies in the field of view.
Our final sample contains 1105 galaxies, of which 538  have S$\rm \acute{e}$rsic indices $n < 2.5$ (disk-like) and 567  have  $n \ge 2.5$ (elliptical-like). 
Note that the S$\rm \acute{e}$rsic index $n$ is not an ideal indicator of galaxy morphology: some disk galaxies may 
have $n\ge2.5$ while some elliptical  galaxies may have $n<2.5$. 
Broad band colours, such as $NUV-r$ \citep[e.g.][]{schiminovich07}, may help separate passive and star-forming galaxies 
and we will analyze the impact of the $NUV-r$ colour in the following sections.

\section{Method}
\label{methods}
We measure the density field around each MaNGA galaxy and classify its environment into one of four categories: cluster, filament, sheet, or void. We  measure its stellar age and metallicity at different parts of the galaxy using full spectral fitting method and then measure gradients and zero points of the age and metallicity radial profiles. Finally we compare the measured the gradients and zero-points of galaxies in different environments. We discuss each of these in turn. 

\subsection{Environment}
\label{envmethod}
In this paper, we consider two environment indicators:  the local mass density and the type of large scale structure (LSS). 
The environment data are taken from the ELUCID project \citep{wang16}, which aims to perform constrained 
simulations of the local universe that can provide an optimal way to utilize and explain the abundant observational data. 
This project uses the halo-domain method developed by \citet{wang09} to reconstruct the cosmic density field 
in the local universe from the SDSS DR7 galaxy group catalogue \citep{yang07}. As shown in \citet{wang16}, 
the reconstructed density field matches well the distributions of both the galaxies and groups. 
The local mass density environment indicator of a galaxy is the density at the position of the galaxy smoothed by
a Gaussian kernel with a smoothing scale of $1 Mpc/h$.  Since the density field is reconstructed in real space, we have to correct
for redshift distortions in the distribution of galaxies. The Kasier effect is corrected by using the method developed 
in \citet{wang12}. To correct for the {\it finger of god} effect, we first crossmatch our galaxy sample with the galaxy catalogue 
used to construct the group catalogue. We then use the group centre to represent the galaxy position in deriving the 
density field. We refer the reader to \citet{wang16} for the details of the reconstruction methods.

The morphology of the LSS is very complex. Recently, much effort has been given to developing methods 
to classify the cosmic web \cite[e.g.][]{hahn07,hoffman12}. Here we adopt a dynamical classification method
developed by \citet{hahn07}, which uses the eigenvalues of the tidal tensor to determine the type of the local 
structure in a cosmic web. The tidal tensor, ${\cal T}_{ij}$, is defined as
\begin{equation}\label{eq_tij}
{\cal T}_{ij}=\partial_i\partial_j\phi\,,
\end{equation}
where, $\phi$ is the peculiar gravitational potential and can be
calculated from the reconstructed density field shown above. Following \citet{hahn07}, we smoothed the density field with a Gaussian kernel with a smoothing scale of 2Mpc/h. We then diagonalize the tidal field tensor $T_{ij}$ and derive the eigenvalues $T_1$, $T_2$, and $T_3$, which corresponds to the major, intermediate and minor axes of the tidal field. The LSS environment is classified into four categories following the definition by \citet{hahn07}: $cluster$ has three positive eigenvalues ($T_1 > 0, \,T_2 > 0,\, T_3 >0$, fixed points); $filament$ has two positive and one negative eigenvalues ($T_1 > 0,\, T_2 > 0, \,T_3 < 0$, two-dimensional stable manifold); $sheet$ has one positive and two negative eigenvalues ($T_1 > 0, \,T_2 < 0,\, T_3 < 0$, one-dimensional stable manifold); and $void$ has three negative eigenvalues ($T_1 < 0, \,T_2 < 0, \,T_3 < 0$, unstable orbits). The method has been shown to very effectively classify the large scale structures of local Universe \citep[see][]{wang16}.

\subsection{Stellar population analysis}
\label{spsmodel}
We bin  the MaNGA datacube of a galaxy using the Voronoi binning method \citep{cc03} so that each binned spectrum has a signal-to-noise ratio (SNR) equal to $\sim 20$ (determined in the 5800-6400\AA wavelength range). Spaxels with $\rm SNR \le 5$ or dominated by sky lines are removed before binning. We fit the Voronoi binned spectra using the STARLIGHT code \citep{cf05} using 150 bases from the model of \citet{bc03} with the STELIB stellar library \citep{lb03}  and using a \citet{chabrier03} IMF to obtain the stellar population parameters. 
The STARLIGHT bases are single stellar population (SSP) templates defined using a grid of ages and metallicities. There are 25 ages (0.001, 0.003, 0.005, 0.007, 0.009, 0.01, 0.014, 0.025, 0.04, 0.055, 0.102, 0.161, 0.286, 0.509, 0.905, 1.278, 1.434, 2.5, 4.25, 6.25, 7.5, 10, 13, 15, 18 Gyr) and 6 metallicities (0.0001, 0.0004, 0.004, 0.008, 0.02, 0.05). 
We use the 3710-8000\AA \,part of the de-redshifted spectra in the STARLIGHT fitting because the parts of the spectrum outside this range are sometimes dominated by noise. 
Emission line regions are masked out using the STARLIGHT standard emission line mask file and other regions are weighted equally during the fitting. Note that in our companion paper by Goddard et al. (submitted),  the FIREFLY software \citep{wilkinson15} is used with the \citet{ms11} model utilising the  MILES stellar library \citep{sb06} and a \citet{kroupa01} IMF. In addition, they have a slightly different Voronoi binning scheme and their fitting wavelength range is 3600-6900\AA. (This different choice of wavelength range is 
owing to the wavelength restriction in their models.) Our results are in good agreement despite these differences. 

The STARLIGHT code returns the best-fitting model, which is a mixture of different bases. Both the light fraction (determined around wavelength 4020 \AA\footnote{The choice of 4020\,\AA\, is the default setting of STARLIGHT. This wavelength range does not have emission lines and it has been proven to be a good choice by extensive tests.}) $x$ and the current mass fraction $M$ of each base are listed in the output file. We calculate the luminosity-weighted age using
\begin{equation}
{\tau_L} =\frac {\Sigma_{j} x_j \tau_j} {\Sigma_{j} x_j},
\end{equation}
where $\tau_j$ is the age value of the $j$-th base and $x_j$ is the light fraction of the $j$-th base. Similarly, we calculate the luminosity-weighted metallicity using
\begin{equation}
Z_L = \frac {\Sigma_j x_j Z_j} {\Sigma_j x_j},
\end{equation}
where $Z_j$ is the metallicity value of the $j$-th base. We calculate the mass-weighted age and metallicity using
\begin{equation}
{\tau_M} =\frac {\Sigma_{j} M_{j} \tau_j} {\Sigma_{j} M_{j}},
\end{equation}
and
\begin{equation}
Z_M =\frac {\Sigma_{j} M_{j} Z_j} {\Sigma_{j} M_{j}},
\end{equation}
where $M_{j}$ is the current mass fraction of the $j$-th base. Note that this is different from some earlier studies, e.g. \citet{gd15}, who use log($\tau_j$) and log($Z_j$) before weighting. We use this definition because it is physically more intuitive. 
Also, the weighting formalism used by \citet{gd15} may give more weight 
to younger and metal poorer stellar populations.

After fitting the whole datacube, we create maps of the stellar population parameters, such as stellar age and metallicity.  We radially bin these maps into elliptical annuli with widths of 0.1 effective radius ($R_e$).  Here $R_e$ is defined as the major axis of the ellipse that 
contains 50\% of the total $r$-band flux of the galaxy. The effective radius, position angle and ellipticity of the galaxy are obtained by fitting Multi-Gaussian Expansion (MGE) models \citep{cappellari13} to the galaxy's $r$-band image. Finally, we derive the stellar population radial profiles using the median values of each annulus.

\begin{figure} 
\begin{center}
\includegraphics[scale=0.4]{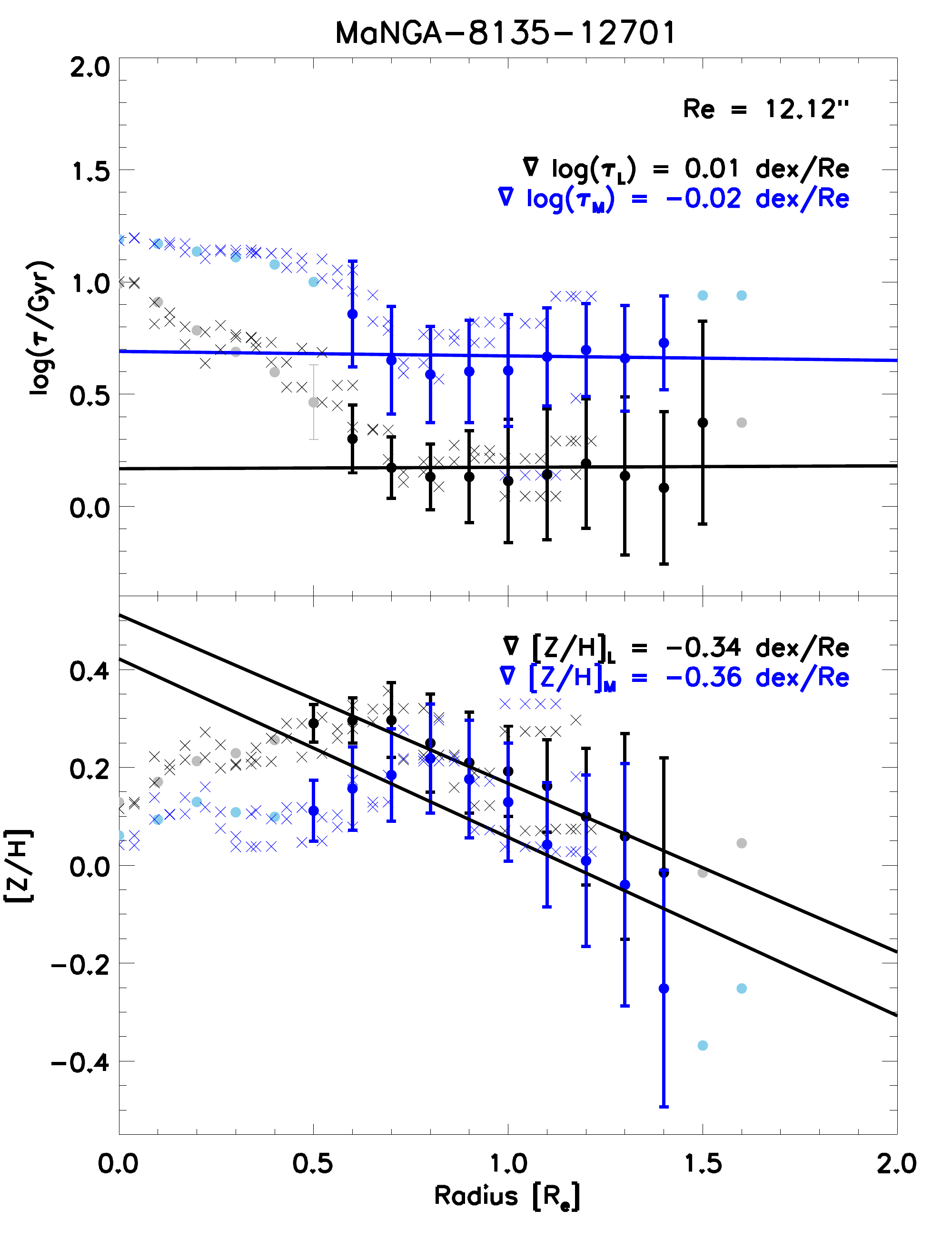}  

\caption{A line fitting example for a MaNGA galaxy (the galaxy shown in the first row of Fig. \ref{2dmaps}.) Upper panel: linear fitting of  age profiles; lower panel: linear fitting of metallicity profiles.  The dots show the median radial profiles and the lines show straight line fittings of the points within $0.5 - 1.5 R_e$. The error bars show $1\sigma$ scatter within each radial bin. The crosses show values along the major axis of this galaxy. Black symbols are for luminosity-weighted values and blue symbols are for mass-weighted values. 
}
\label{fittingexample}
\end{center}
\end{figure}

\section{Results}
\label{results}

\begin{figure} 
\begin{center}
\includegraphics[scale=0.35,angle=90]{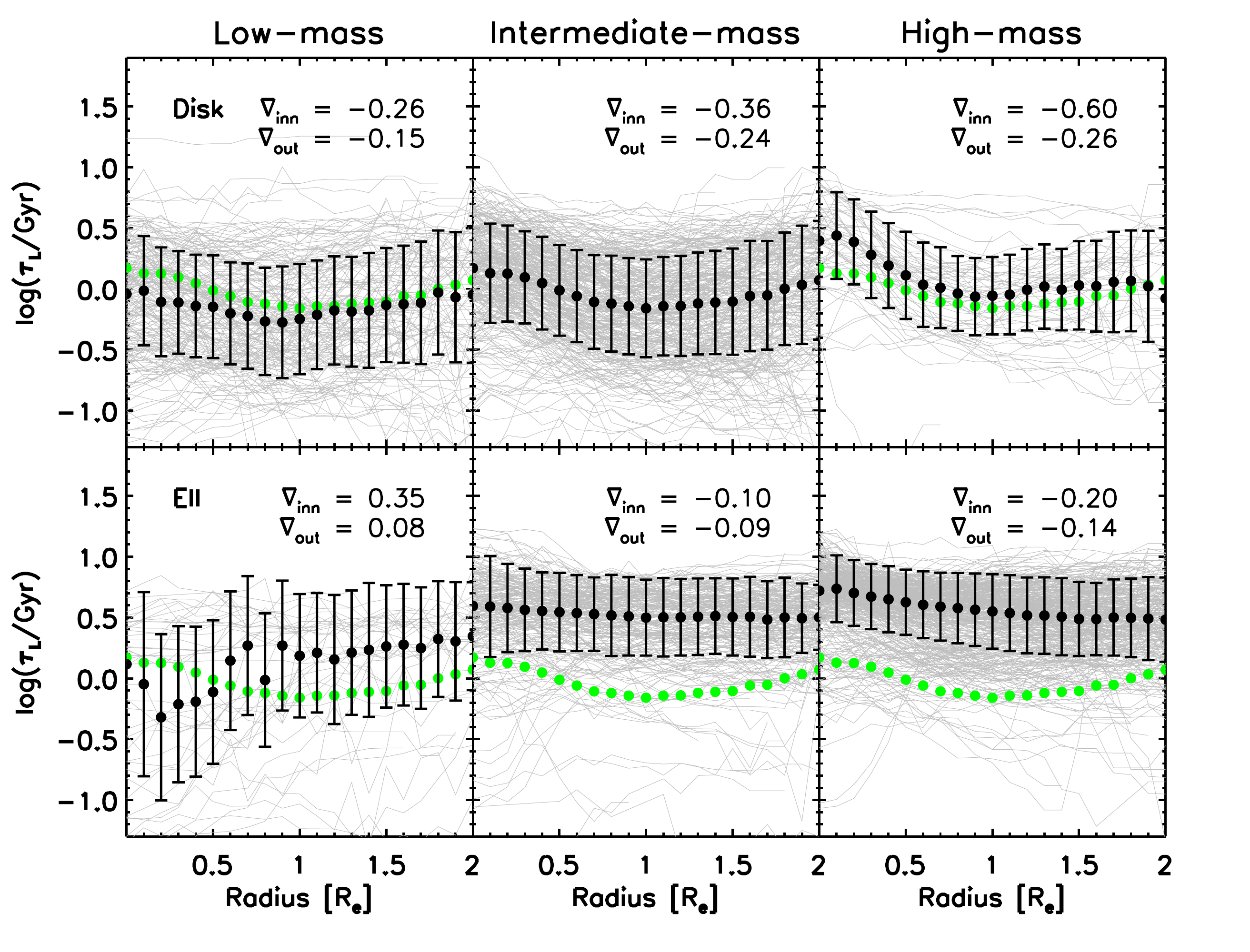}  
\vspace*{-0.2cm }
\caption{Co-added luminosity-weighted age profiles. The galaxies are split into 6 groups based on mass and morphology. Low mass (left column) is $\log(M_*/M_{\odot}) \le 9.5$, intermediate mass (middle column) is $9.5<\log(M_*/M_{\odot}) \le 10.5$, high mass (right column) is $\log(M_*/M_{\odot})>10.5$; disk (top row) is $n<2.5$, and elliptical (bottom row) is $n\ge2.5$.
 The radial profiles are plotted with gray lines. The black dots with error bars show the median value and 1$\sigma$ dispersion of the black lines; the green dots show the median value of intermediate-mass disk galaxies and are the same in all panels. Each panel also shows the gradient of the median profile in the inner region (within $0-1\,R_e$) and in the outer region (within $0.5-1.5\,R_e$)
}
\label{agel_prof}
\end{center}
\end{figure}

\begin{figure} 
\begin{center}
\vspace*{-0.2cm }
\includegraphics[scale=0.35,angle=90]{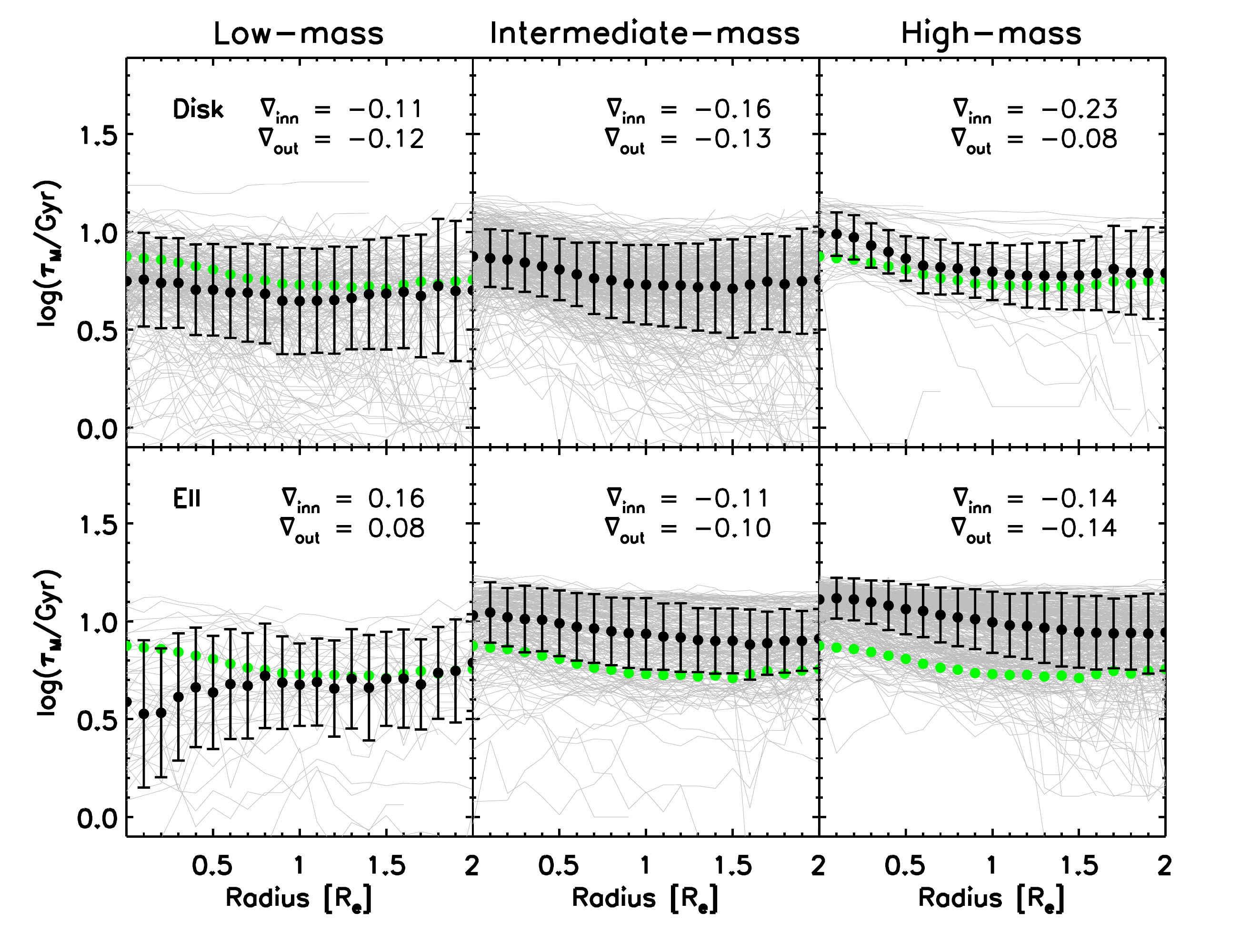}  

\caption{Co-added mass-weighted age profiles. The symbols are the same as in Fig. \ref{agel_prof}.
}
\label{agem_prof}
\end{center}
\end{figure}

\begin{figure} 
\begin{center}
\includegraphics[scale=0.35,angle=90]{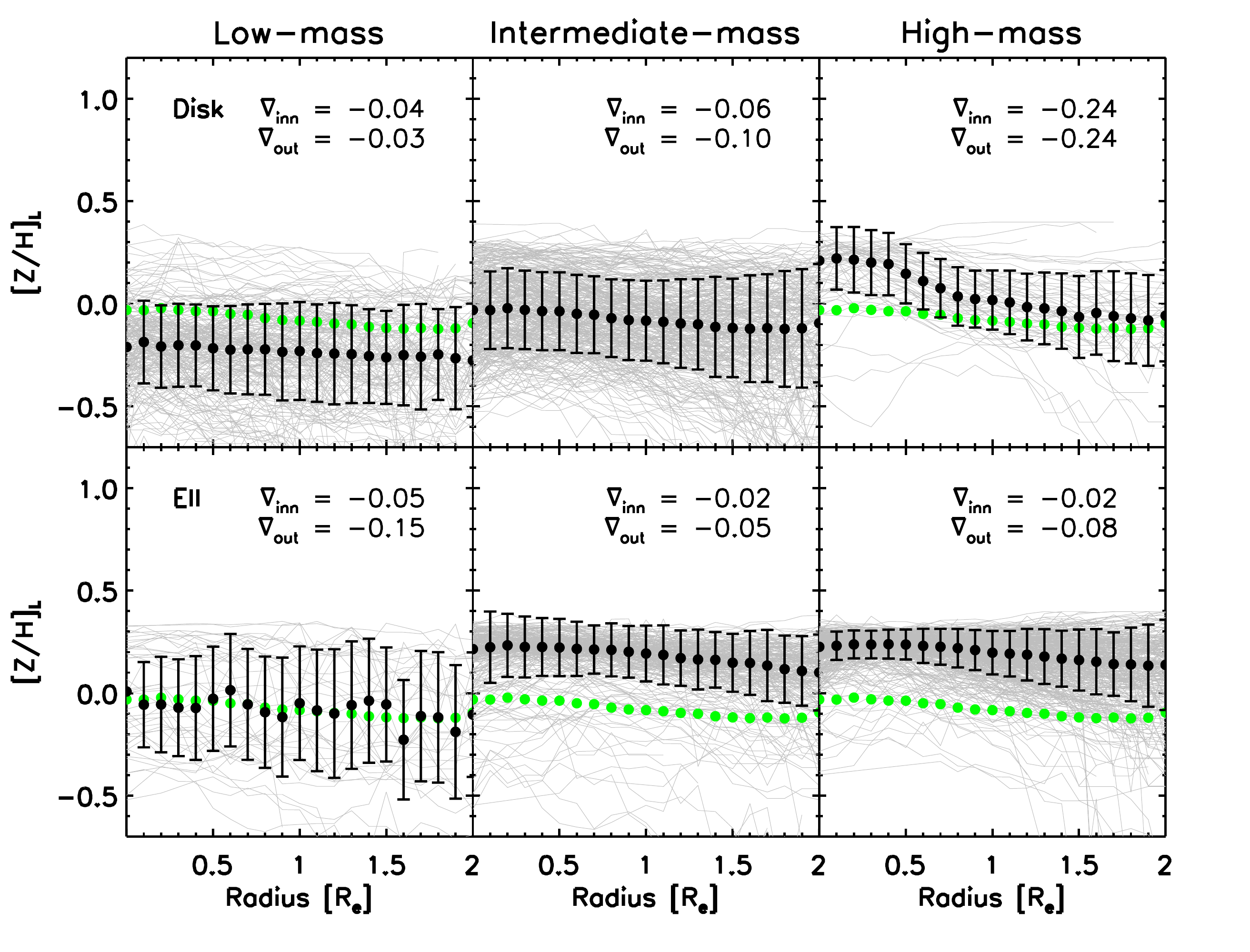}  
\vspace*{-0.2cm }
\caption{Co-added luminosity-weighted metallicity profiles. The symbols are the same as in Fig. \ref{agel_prof}.
}
\label{mzl_prof}
\end{center}
\end{figure}

\begin{figure} 
\begin{center}
\includegraphics[scale=0.35,angle=90]{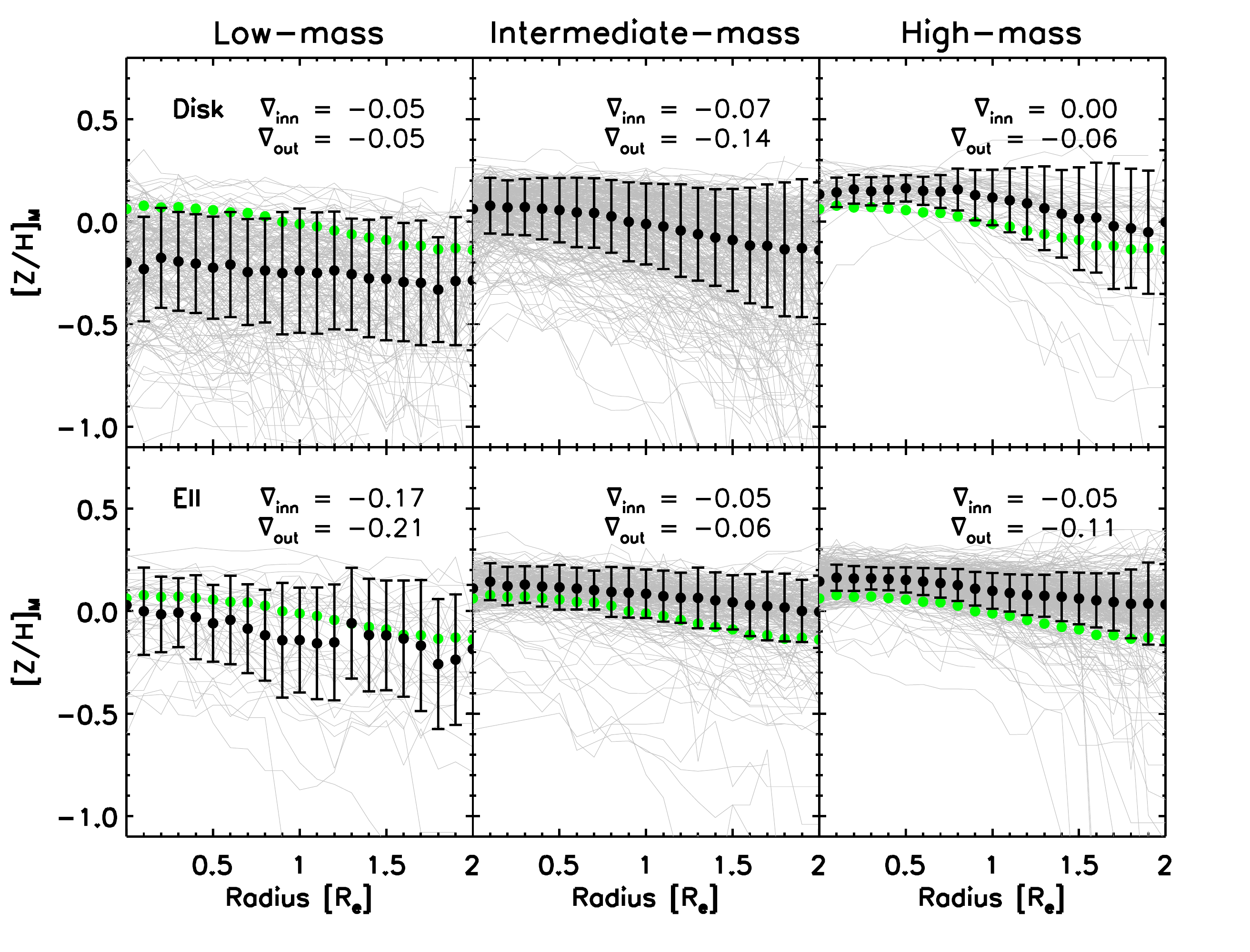}  
\vspace*{-0.2cm }
\caption{Co-added mass-weighted metallicity profiles. The symbols are the same as in Fig. \ref{agel_prof}.
}
\label{mzm_prof}
\end{center}
\end{figure}

In this paper we focus on  the distributions of  stellar age and metallicity because they are the two most important stellar population parameters. We use logarithmic values of age ($\log(\tau)$) and metallicity ([Z/H]) in the remainder of this paper. Five example galaxies observed with different IFU bundle sizes are shown in Fig. \ref{2dmaps}. We can see from these plots that distributions of stellar population parameters could roughly be fitted by concentric elliptical contours. Also, galaxy centres are generally older and more metal rich than the outer regions.

Bulges usually have a steep increase in age and metallicity towards the centre of a galaxy and they may have different formation histories from disks \citep{spiniello15}. 
We focus on the ``bulk'' part of the galaxy, i.e. the more extended disk part. 
Therefore, we  exclude bulges  in studying the relationship between age and metallicity gradients and environments. 
According to \citet{zheng15}, disk galaxies usually have bulge sizes about $0.3\,r_{90}$ ($r_{90}$ 
being the radius that contains 90\% of the total $r$-band flux) or $0.5$ times the effective radius $R_{e}$. 
Therefore, we measure the gradients between $0.5\,R_{e}$ and $1.5\,R_{e}$ because all MaNGA galaxies are fully covered by the IFUs up to $1.5\,R_{e}$.  We have examined the effect on our results if bulges are included and find that there is no significant change to our conclusions except  that we would have larger scatters in the gradients. 
We fit the data points within our radial region with a straight line,  defined as 
\begin{equation}
\label{graddef1}
 \log(\tau(R/R_e)) = \log(\tau(0)) + k_{\tau} \,R/R_e,
\end{equation}
for  the stellar age profile, and 
\begin{equation}
\label{graddef2}
[Z/H](R/R_e) = [Z/H](0) + k_Z \,R/R_e,
\end{equation}
for the stellar metallicity profile, where $R/R_e$ is the radius in units of the effective radius $R_e$, and $k_{\tau}$ and $k_Z$ are age and metallicity gradients in units of dex/$R_e$. Line fitting is performed using the robust linear regression method implemented in the IDL procedure ROBUST\_LINEFIT, which can identify and remove bad data during the fitting.  One example of the fitting is shown in Fig. \ref{fittingexample}. The gradient uncertainty is estimated using a Monte Carlo method, and is typically $\sim 0.1$ dex/$R_e$.

We present both the  gradients and the fitted values at the effective radii, and explore their correlations with different galaxy properties and environments in the following  sections.

\subsection{Radial profiles}
\label{radprof}

Before discussing the profile fitting results, we show the overall behavior of the age and metallicity radial profiles by plotting the co-added radial profiles (Fig. \ref{agel_prof} - \ref{mzm_prof}). We classify our galaxies using stellar mass $M_*$ and S$\rm \acute{e}$rsic index $n$. We divide our sample into three mass bins: low-mass ($\log(M_*/M_{\odot})  \le  9.5$), intermediate-mass ($9.5<\log(M_*/M_{\odot})  \le  10.5$) and high-mass ($\log(M_*/M_{\odot}) > 10.5$). We further divide the galaxy sample into more disk-like ($n<2.5$) and more elliptical-like ($n\ge2.5$) galaxies. In each figure, we show the radial profiles for these different kinds of galaxies in black  and the median value of these radial profiles in red dots. In order to guide the eye, we also plot the radial profiles of intermediate-mass disk galaxies in gray  and their median values in green dots in each panel.

In general, the mass-weighted age profiles are very flat. The luminosity-weighted age profiles are less flat and sometimes show a  `U' shape curve with the minimum value located around $1-1.5\,R_e$.
The median luminosity-weighted age profiles of the elliptical galaxies are about 0.1 dex older than those of the disk galaxies, and there is almost no dependence on $M_*$.  The switch over in the low-mass elliptical panel (lower-left panel) of Fig. \ref{agel_prof} may be due to small sample size. {The individual galaxies with down-turn age profiles towards the centre of the galaxies might be caused by recent star formation (Ge et al. in prep.; Lin et al. in prep.)}

Most metallicity profiles have a decreasing trend with increasing radius. For disk galaxies, metallicity decreases more rapidly with increasing radius than that for more massive galaxies. 
As a result, high-mass disk galaxies have a steeper  (more negative) gradient and low-mass disk galaxies have a shallower (close to zero) gradient.
Elliptical galaxies are generally more metal rich than disk galaxies, but their gradients  look similar at all three mass bins. 

Age gradients are usually steeper in the central region ($0<R<R_e$) than in the outer region 
($0.5R_e<R<1.5R_e$), with the exception that the median luminosity-weighted age profile for 
low-mass ellipticals has a more positive gradient in the centre than in the outer region. 
However, metallicity profiles in the central regions are usually shallower than in the outer regions.

We note that the radial profiles of individual galaxies could deviate from the general behaviors described above.

\subsection{Dependence on the large scale structure (LSS) environment}

\begin{figure} 
\begin{center}
\includegraphics[scale=0.4]{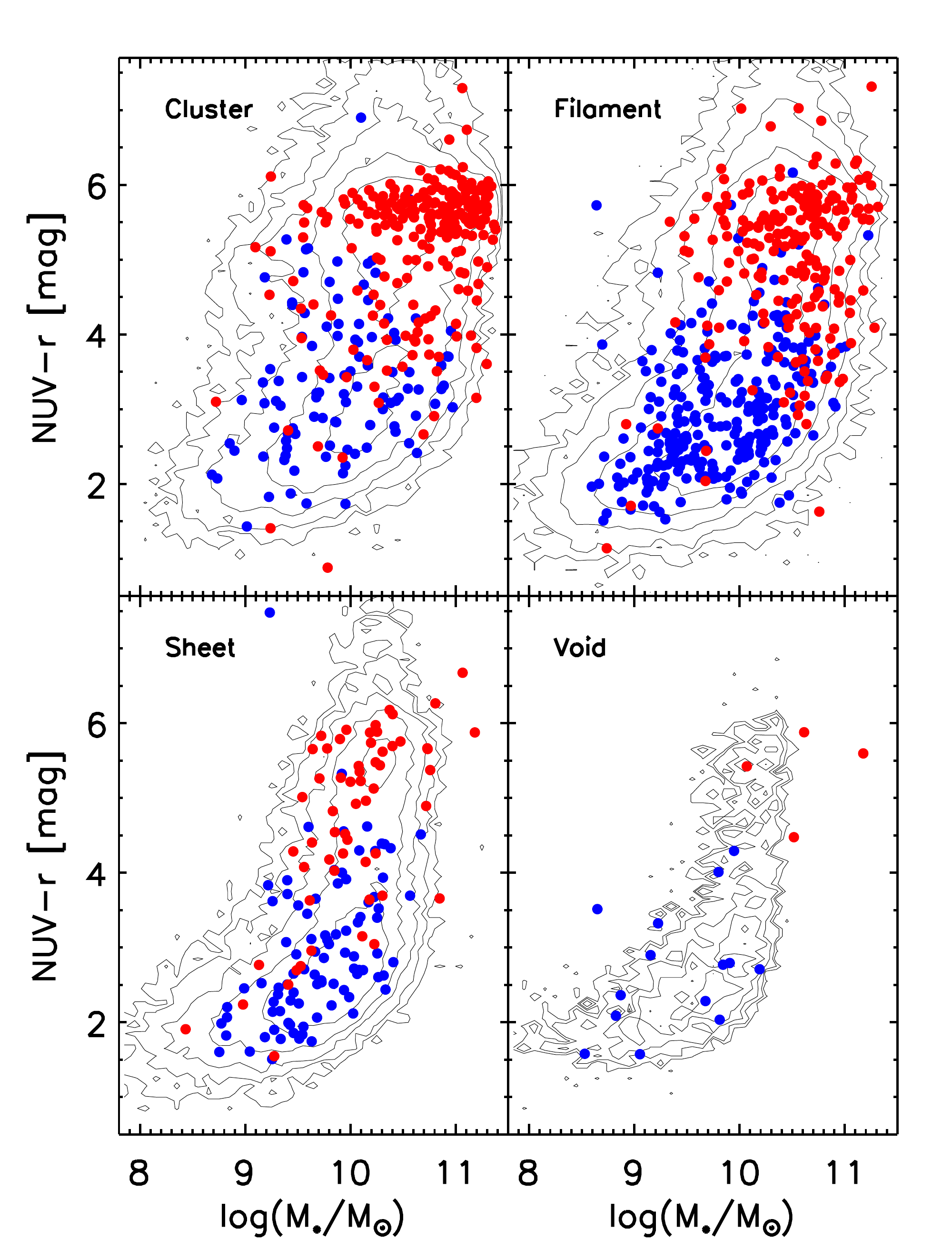}  

\caption{Mass-colour distribution of our sample galaxies in the four LSS environments: 
cluster (upper-left), filament (upper-right), sheet (lower-right), 
and void (lower-right).
Blue dots are for galaxies with S$\rm \acute{e}$rsic indices $n < 2.5$ (disk),
and red dots are for galaxies with S$\rm \acute{e}$rsic indices $n \ge 2.5$ (elliptical). 
Black contours in the background show the distributions of SDSS DR7 galaxies
in the four different LSS environments.}
\label{galpropvsenv}
\end{center}
\end{figure}

The LSS environment dependence is the focus of our study. 
Since stellar mass and colour are known to be correlated with environments \citep{kauffmann04, bm09}, we need to include these parameters in the study. 
Thus, before showing the LSS environmental dependence, we plot the distributions of two important overall galaxy properties, 
i.e. stellar mass $M_*$ and $NUV-r$ colour, in different LSS environments (Fig. \ref{galpropvsenv}). Note that red galaxies have a much larger portion in denser environments and this is consistent with \citet{ll08, bm09,thomas10}

In the following two subsections we explore the LSS environment dependence of both gradients and zero points,
i.e. the fitted values at the effective radius $R_e$. 

\subsubsection{Gradients versus LSS environment}

\begin{figure*} 
\begin{center}
\includegraphics[scale=0.45,angle=90]{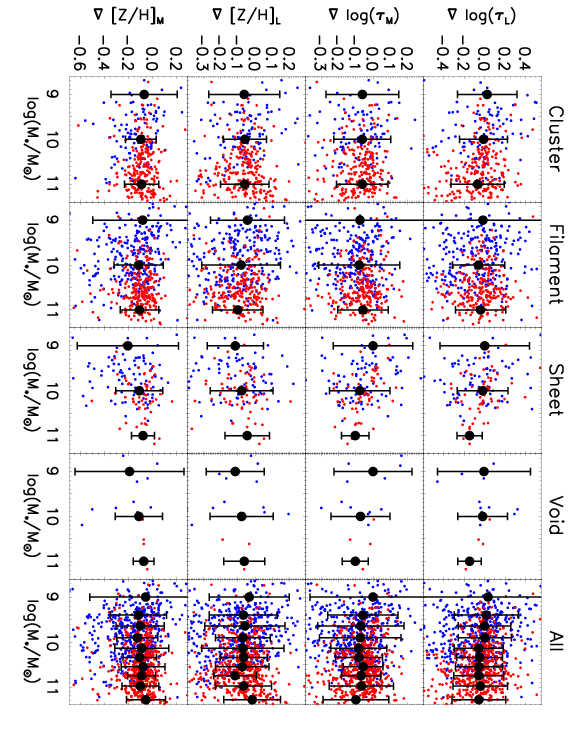}  

\caption{Age and metallicity gradients in units of dex/$R_e$ versus  total stellar mass $M_*$ for galaxies in different environments.  Red dots are for elliptical galaxies ($n \ge 2.5$) and blue dots are for disk galaxies ($n < 2.5$). The big black dots with error bars show the median value and 1$\sigma$ scatter of gradients in the three mass bins: $\log(M_*/M_{\odot}) \le 9.5$, $9.5<\log(M_*/M_{\odot}) \le 10.5$, and $\log(M_*/M_{\odot})>10.5$. The right column show gradients of all galaxies in our sample using finer mass bins to explore trends with $M_*$. 
}
\label{gradvsm}
\end{center}
\end{figure*}

\begin{figure*} 
\begin{center}
\includegraphics[scale=0.45,angle=90]{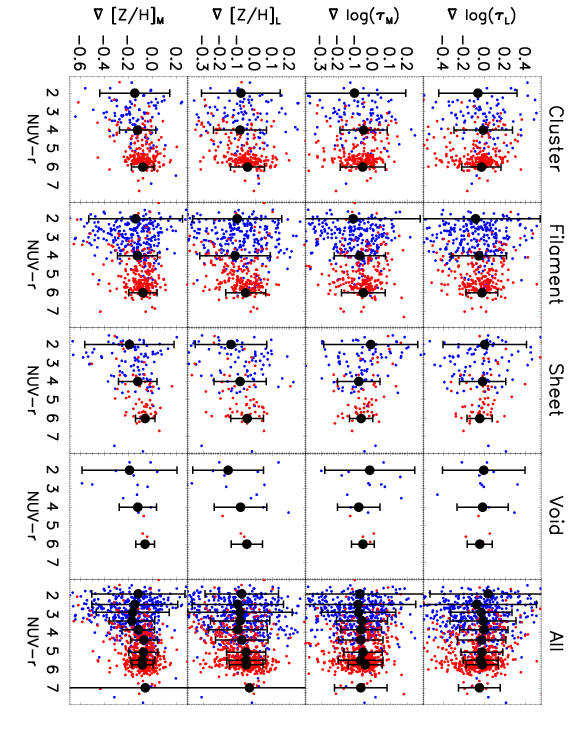}  

\caption{Age and metallicity gradients in units of dex/$R_e$ versus $NUV-r$ for galaxies in different environments. The right column show gradients of all galaxies in our sample. Red dots are for elliptical galaxies ($n \ge 2.5$) and blue dots are for disk galaxies ($n < 2.5$). The big black dots with error bars show the median value and $1\sigma$ scatter of gradients in the three colour bins: $NUV-r \le 3$, $3<NUV-r\le 5$, and $NUV-r>5$. The right column show gradients of all galaxies in our sample and using finer colour bins to explore trends with $NUV-r$ colour. 
}
\label{gradvsnuvr}
\end{center}
\end{figure*}

We first examine the correlation between the stellar population gradients and the LSS environments. 
Here we plot the age and metallicity gradients versus $M_*$, (Fig. \ref{gradvsm}) and $NUV-r$ colour (Fig. \ref{gradvsnuvr}) for different environments.  
Both age and metallicity gradients are close to zero and slightly negative. 
They appear to correlate weakly with $M_*$ and $NUV-r$ colour although the correlations changes in different panels of the plots and the scatters are large.
The distributions of gradients appear slightly different in different LSS environments, but the difference could be dominated by $M_*$, although the scatters are large for these plots. 
We also show gradients of all the galaxies in our sample in the rightest columns of Figs. \ref{gradvsm} and \ref{gradvsnuvr}. 

\begin{figure} 
\begin{center}
\includegraphics[scale=0.35]{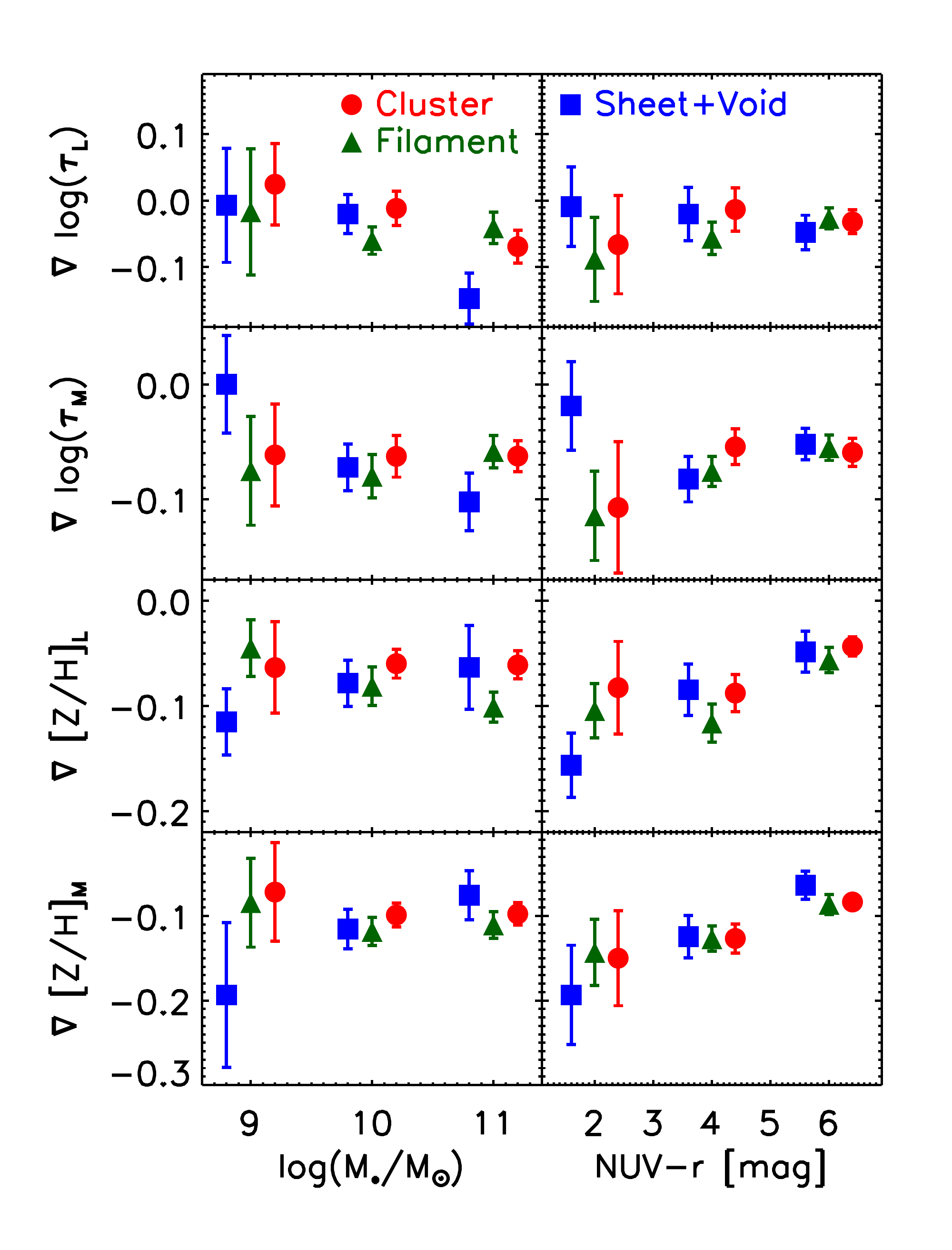}  

\caption{Median gradients and their errors in each mass bin (left panels) and $NUV-r$ bin (right panels). The x-axes are shifted a little bit so that the points do not overlap with each other in the figure. The  three mass bins are $\log(M_*/M_{\odot}) \le 9.5$, $9.5<\log(M_*/M_{\odot}) \le 10.5$, and $\log(M_*/M_{\odot})>10.5$; and the three $NUV-r$ colour bins are  $NUV-r \le 3$, $3<NUV-r \le 5$, and $NUV-r>5$. Red dots show cluster environment,  dark green triangles show filament environment, and blue squares show sheet and void environment. A careful reader may have noticed that the error bars in this figure are smaller than those in Fig. \ref{gradvsm} and \ref{gradvsnuvr}. This is because the error bars shown here are errors of the median values, which are $1.253\,\sigma/\sqrt{N_g}$, where $N_g$ is the number of galaxies in each mass or colour bin.} 
\label{gradvsenv}
\end{center}
\end{figure}

In order to examine the environmental dependence more clearly, we plot the median gradients and the errors of the median values (instead of $1\sigma$ scatter) in each mass bin and $NUV-r$ colour bin in Fig. \ref{gradvsenv}. 
For low-mass and blue galaxies, there is slight environmental dependence, but differences in different environments are mostly within uncertainties.\footnote{Note different bins have different number of galaxies and small error bar is usually dominated by large galaxy number in that bin.} For high-mass and red galaxies, there are nearly no dependences on LSS environments. 

\begin{figure*} 
\begin{center}
\includegraphics[scale=0.5,angle=90]{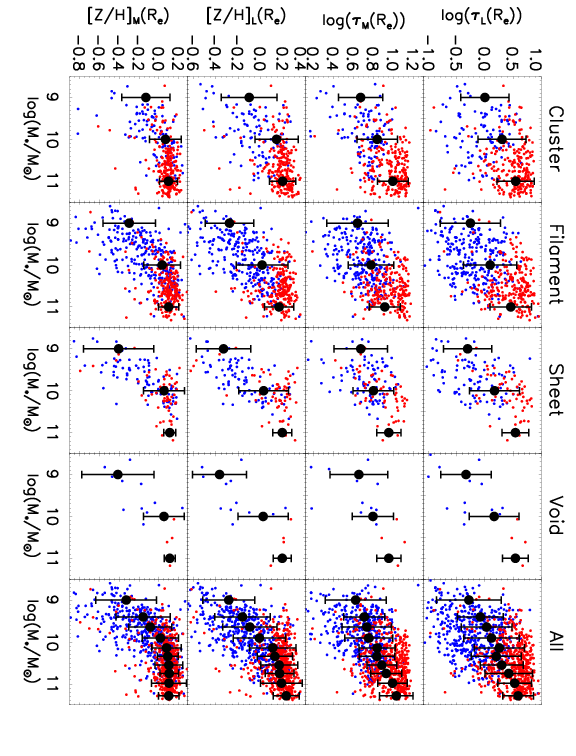}  
\vspace*{-0.5cm }
\caption{ The fitted age and metallicity values at the effective radius $R_e$. Symbols are the same as Fig. \ref{gradvsm}. 
}
\label{zeropm}
\end{center}
\end{figure*}

\subsubsection{Values at $R_e$ versus LSS environment}

\label{vrelss}

\begin{figure*} 
\begin{center}
\includegraphics[scale=0.5,angle=90]{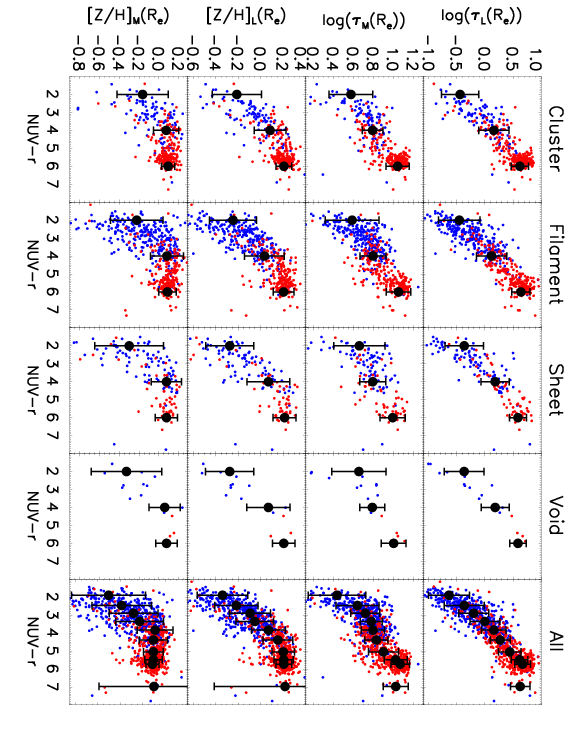}  
\vspace*{-0.5cm }
\caption{ The fitted age and metallicity values at the effective radius $R_e$. Symbols are the same as Fig. \ref{gradvsnuvr}. 
}
\label{zeropnuvr}
\end{center}
\end{figure*}

\begin{figure} 
\begin{center}
\includegraphics[scale=0.35]{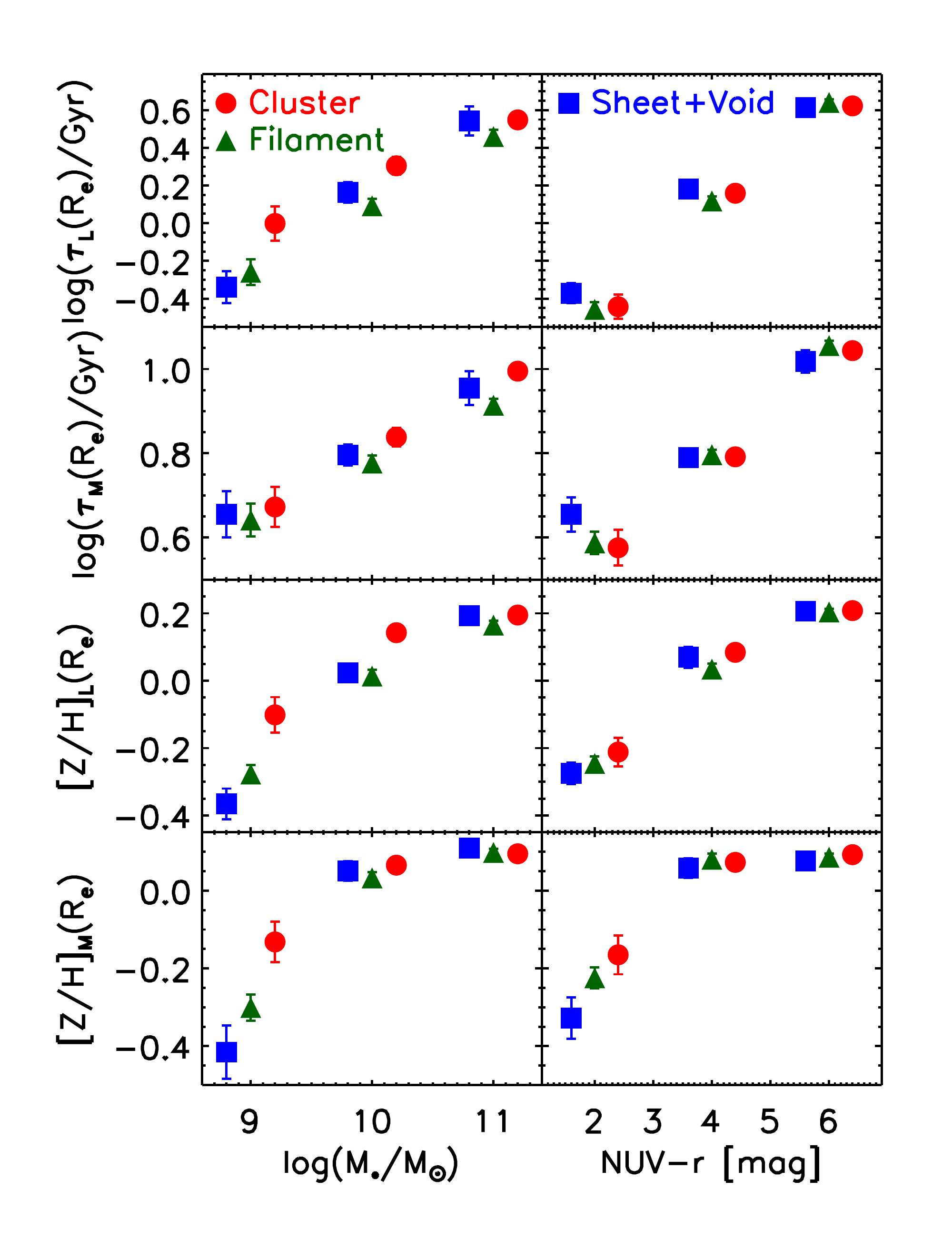}  
\vspace*{-0.3cm }
\caption{Median values of fitted age and metallicity at the effective radius $R_e$ in each mass (left panels) and $NUV-r$ colour (right panels) bin. Symbols are the same as Fig. \ref{gradvsenv}. 
}
\label{zeropvsenv}
\end{center}
\end{figure}

An important and also complementary quantity from our line fitting is the fitted age and metallicity values at $R_e$. These are similar quantities to the central values derived using spectra from single central fibres \cite[e.g.][]{kauffmann04}. We plot these values in a similar way to Figs. \ref{gradvsm} - \ref{gradvsnuvr} and the results are shown in Figs. \ref{zeropm} - \ref{zeropnuvr}.  
Ages at $R_e$ for both elliptical and disk galaxies are highly correlated with both $M_*$ and $NUV-r$ colour.
The metallicity at $R_e$ is also correlated with $M_*$ and $NUV-r$ colour for disk galaxies. For elliptical galaxies, however, 
[Z/H]($R_e$) have similar values.

The LSS environment dependence is more obvious in Fig. \ref{zeropvsenv}, in which we plot the median values in each mass and colour bin. 
Low-mass and blue galaxies seem to have some correlations with environment: they have younger ages and lower metallicities in low density regions. Intermediate-mass and high-mass galaxies do not appear to show this  trend.  This implies that the LSS environment plays a greater role for smaller galaxies. For massive galaxies, stellar mass plays a dominant role so it matters less where they are located. However, we note that mass-weighted ages of low-mass galaxies and both luminosity and mass-weighted ages of blue galaxies do not show the correlation mentioned above.

\subsection{Dependence on local densities}

The other important environment indicator is the local density. 
We calculate the average local mass density within $1\,$Mpc of each individual galaxy based on the reconstructed density field \citep{wang09, wang14}. This is a similar environment indicator to the one used by our companion paper Goddard et al. (submitted) and some previous studies \cite[e.g.][]{kauffmann04}. 

\begin{figure} 
\begin{center}
\vspace*{-0.5cm }
\includegraphics[scale=0.35]{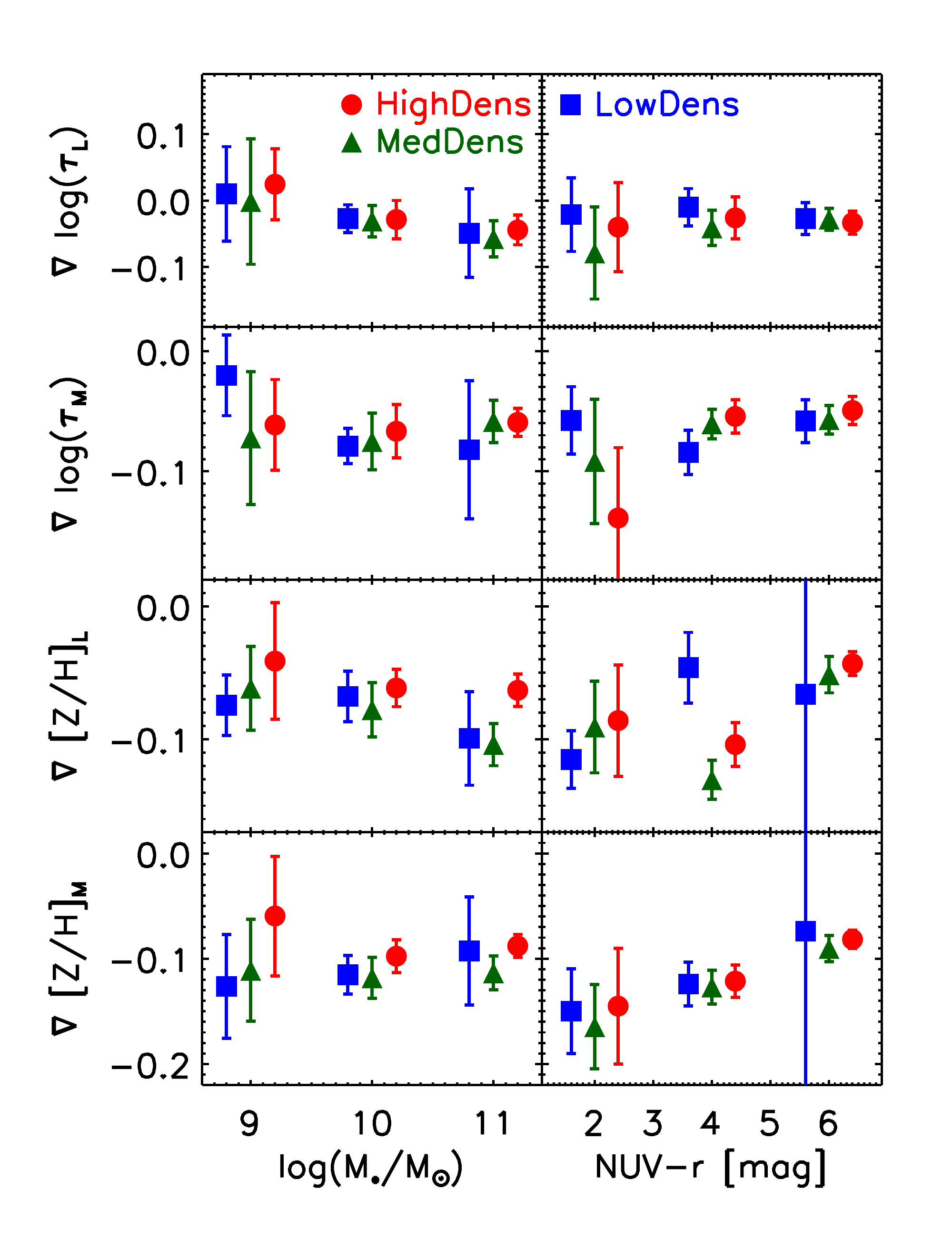}  
\vspace*{-0.3cm }
\caption{Median gradients and their errors in each mass bin (left panels) and $NUV-r$ bin (right panels). The symbols are now colour-coded in local densities: blue squares show galaxies in low-density ($d_l \le 2$) regions, green triangles show galaxies in intermediate-density ($2<d_l \le 10$) regions, and red dots show galaxies in high-density ($d_l>10$) regions, where the local density, $d_l$, is the average mass density within $1\,$Mpc of the target galaxy and is in units of average cosmic mean density, i.e. $7.16\times 10^{10} {M}_{\odot}/h/({Mpc}/h)^3$ assuming WMAP5 cosmology.  The mass-bins and colour-bins are the same as Fig. \ref{gradvsenv}. }
\label{gradvslocalden}
\end{center}
\end{figure}

We present the age and metallicity gradients versus the local density in Fig. \ref{gradvslocalden}. Galaxies are colour-coded in $M_*$, and the mass bins are as defined in Section \ref{radprof}. Similar to the LSS environment results, some gradients show a weak trend along different local density environments. For example, low mass galaxies (top-left panel of Fig. \ref{gradvslocalden}) show an increase of luminosity-weighted age gradient towards dense regions,  however the amplitude of variations is comparable to the errors.

\begin{figure} 
\begin{center}
\includegraphics[scale=0.35]{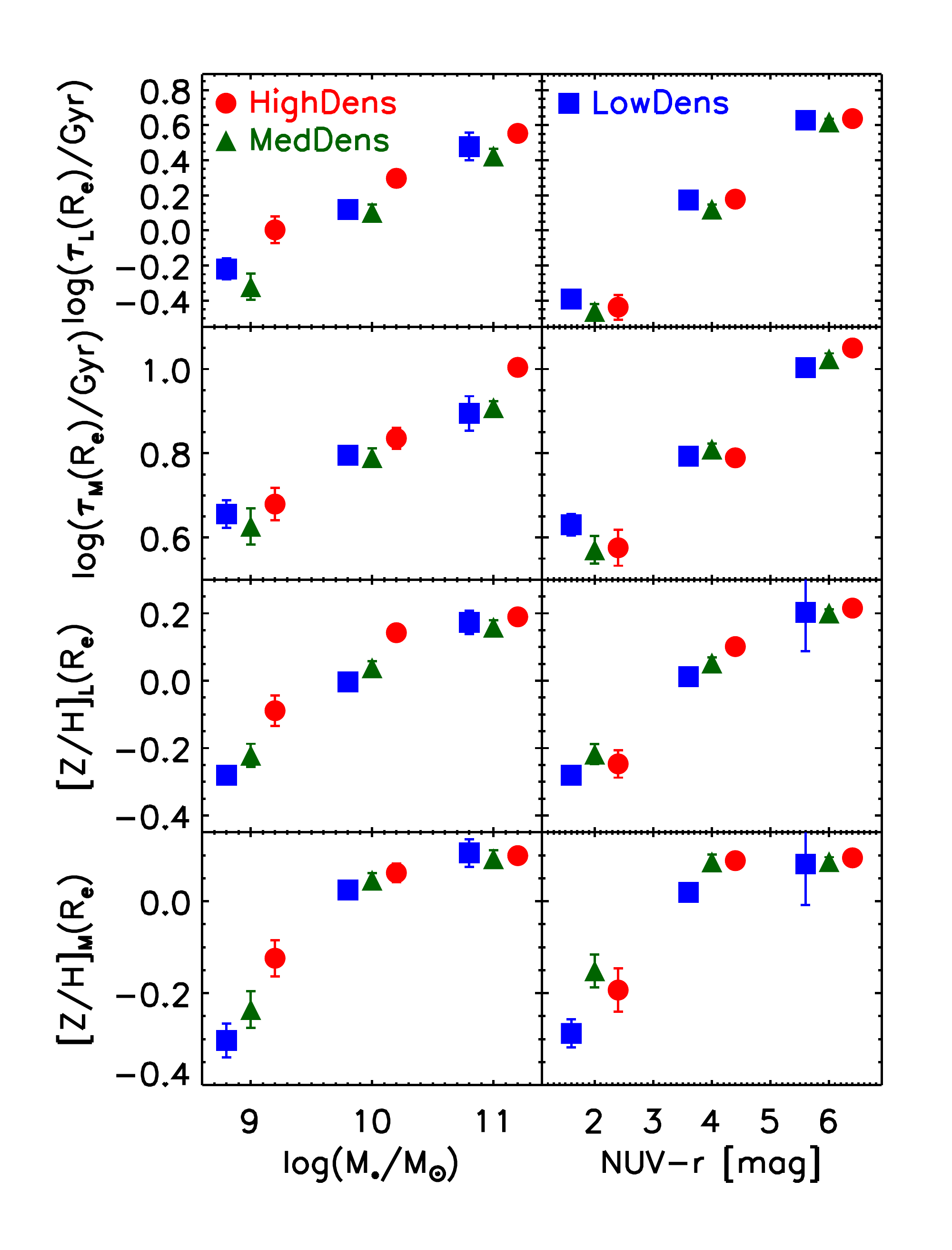}  
\vspace*{-0.3cm }
\caption{Median values of fitted age and metallicity at the effective radius $R_e$ versus local densities.  Symbols are the same as 
Fig. \ref{gradvslocalden}}
\label{zeropvslocalden}
\end{center}
\end{figure}

We also plot the median values of fitted age and metallicity at the effective radius $R_e$ versus local densities in Fig. \ref{zeropvslocalden}. The fitted values at $R_e$ show an obvious trend along different local densities. 
Galaxies have younger ages and lower metallicities in low local density regions. 
This trend is more obvious for low-mass and blue galaxies and is consistent with the result of previous studies \citep[e.g.][]{kauffmann04, lin14}.
Again, as in Section \ref{vrelss}, mass-weighted ages of low-mass galaxies and both luminosity and mass-weighted ages of blue galaxies do not show the correlation mentioned above.

\section{Discussion}
\label{discussion}

\subsection{Comparison with other studies}

\subsubsection{Gradients}
There have been many theoretical predictions for stellar population distributions using different galaxy formation and evolution models. Generally speaking, monolithic collapse models \cite[e.g.][]{pipino10} predict steep metallicity gradients for elliptical galaxies, with a typical slope of about $-0.3 \, \rm dex/dex$. Mergers and stellar migrations would make this gradient much flatter \citep{roskar08, sb09, dm09, minchev12}.

There have also been many observational studies on age and metallicity gradients for both disk and elliptical galaxies. Almost all of them found slightly negative gradients \citep{roig15, morelli15, sb14, gd15}, which is consistent with the inside-out galaxy formation scenario. Many previous analyses used a logarithmic definition of gradients, defined as 
\begin{equation}
\label{graddef3}
 \log(\tau(\log (R/R_e))) = \log(\tau(0)) + k_{\tau,lg} \,\log(R/R_e),
\end{equation}
for  the stellar age profile, and 
\begin{equation}
\label{graddef4}
[Z/H](\log(R/R_e)) = [Z/H](0) + k_{Z,lg} \,\log(R/R_e).
\end{equation}
We have also computed gradients using this definition to better compare with previous results. Note 
that this definition works better for larger radial ranges and gives more weight to the inner regions. 
We compile the gradients estimated by several recent studies as well as our own results in Table \ref{compare}.  Results calculated using eqs. \ref{graddef1}-\ref{graddef2} have units of dex$/R_e$ 
and results calculated using eqs.  \ref{graddef3}-\ref{graddef4} have units of dex/dex. 

\citet{rawle10} derived stellar population distributions for 25 early type cluster galaxies using spectral index information. They found that the mean stellar metallicity gradient for their sample galaxies is $-0.13 \pm 0.04$ dex/dex and the mean age gradient is $-0.02\pm0.06$ dex/dex. \citet{kuntschner10} using a similar spectral index analysis method, studied 48 early type galaxies from the SAURON\footnote{Spectroscopic Areal Unit for Research on Optical Nebulae \citep{bacon01}} sample and found a mean metallicity gradient value of $-0.25\pm0.11$ dex/dex for old ellipticals and $-0.28\pm0.12$ dex/dex for young ellipticals, and a mean age gradient value of $0.02\pm0.13$ dex/dex for old galaxies and $0.28\pm0.16$ dex/dex for young ellipticals. 

\citet{sb14} using the full spectral fitting method studied 62 disk galaxies from the CALIFA\footnote{Calar Alto Legacy Integral Field Area \citep{sanchez12}} sample and found $\triangledown [Z/H]_M = 0.000\pm0.006$ dex/$R_e$ and $\triangledown \tau_M = -0.087\pm0.008$ dex/$R_e$. They further claim that there is no metallicity gradient - $M_*$ correlation.

More recently, \citet{gd15} studied about 300 CALIFA galaxies with various morphologies and found $\triangledown [Z/H]_M = -0.1\pm0.15$ dex/dex for the inner regions (within $1\, R_e$) of elliptical and S0 galaxies and a shallower metallicity slope in the 1-3 $R_e$ region. They also found that the metallicity gradients are almost constant for S0 and E (and very massive) galaxies but that there is a weak correlation between metallicity gradient and stellar mass (and morphological types) for disk galaxies. The metallicity gradients become steeper at higher stellar masses
 with the $\triangledown [Z/H]_M$ ranging from -0.4 to 0.4 dex/$R_e$.

In this paper, we found $\triangledown [Z/H]_M= -0.31\pm0.04$ dex/dex (or $-0.14\pm0.02$\, dex/$R_e$) for disk galaxies and $\triangledown [Z/H]_M=  -0.19\pm0.03$ dex/dex (or $-0.09\pm0.01$\, dex/$R_e$) for elliptical galaxies. 
We also found $\triangledown \tau_M= -0.18\pm0.03$  dex/dex (or $-0.08\pm0.02$\, dex/$R_e$) for disk galaxies and $\triangledown \tau_M=  -0.13\pm0.02$ dex/dex (or $-0.05\pm0.01$\, dex/$R_e$) for elliptical galaxies. Note these mean values and uncertainties are calculated using the MaNGA primary and secondary sample galaxies only because other galaxies do not have a well defined volume weight. We apply a volume weight correction according to the MaNGA sample selection paper Wake et al. (in prep.).
These mean values are consistent with all previous studies.  

We found weak or no correlation between the gradients and $M_*$, which is consistent with \citet{sb14} but slightly different from \citet{gd15}. Our  metallicity gradients are shallower than predictions from most of monolithic collapse models by \citet{pipino10}. 
Many previous studies \citep[e.g.][]{tortora10, gd15}  found that colours and metallicity gradients in spiral galaxies are steeper than in elliptical galaxies. 
\citet{roediger11} also found a similar trend using galaxies in the Virgo cluster.  We also found age and metallicity gradients are steeper in disk galaxies than in ellipticals, consistent with these studies.

\begin{figure} 
\begin{center}
\includegraphics[scale=0.4]{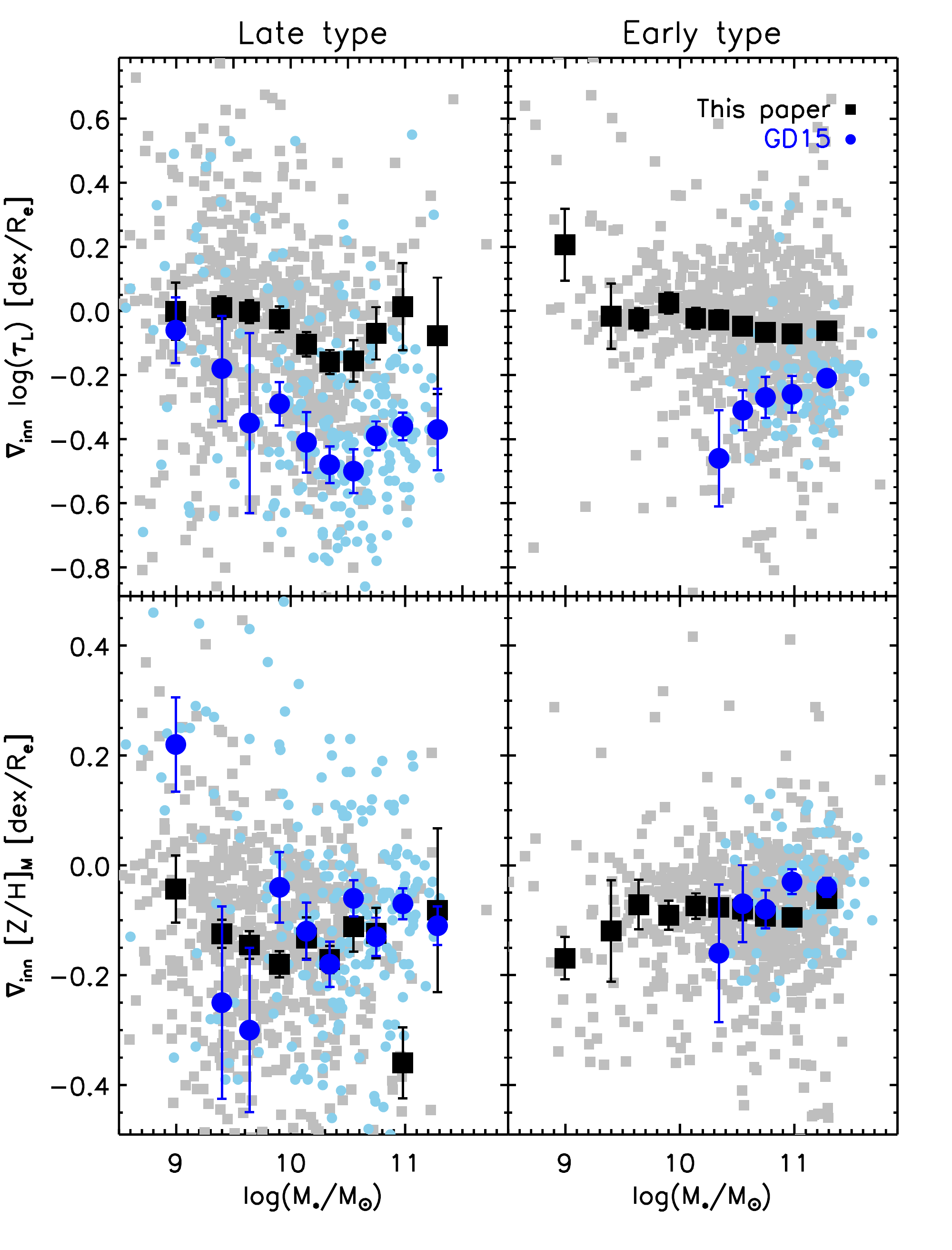}  

\caption{Comparison of inner gradients (between $0-1\,R_e$) between this paper and CALIFA analyses \citep[][using GME models]{gd15}. Upper panels show the luminosity-weighted age gradient and lower panels show the mass-weighted metallicity gradient. Left panels show late-type galaxies and right panels show early type galaxies. Late type is defined as $n<2.5$ for our galaxies and Hubble type $> 0$ for CALIFA galaxies; and early type is defined as $n\ge2.5$ for our galaxies and Hubble type $\le 0$ for CALIFA galaxies. Gray squares are our results for individual galaxies, and sky blue points are CALIFA results for individual galaxies. Big black squares and big blue dots with error bars are the median values of each mass bins with standard errors of the median. 
}
\label{comparecalifa}
\end{center}
\end{figure}

To provide a more direct comparison, we also compute the gradients within $0-1\,R_e$ and plot the 
them versus the total stellar mass on top of CALIFA results \citep{gd15} in Fig. \ref{comparecalifa}. 
The CALIFA results shown here are derived using values at $R_e$ subtracting values at the centre, 
which is slightly different from our linear fitting. The mass-weighted metallicities (lower panels of Fig. \ref{comparecalifa}) 
are consistent with each other. The luminosity-weighted age (top panels of Fig. \ref{comparecalifa}), 
on the other hand,  has about 0.2 dex difference. This difference could be due to their different definition of 
luminosity-weighted age, as their definition tends to put more weight  on younger stellar populations 
(see Section \ref{spsmodel}).

Some previous studies \citep[e.g.][]{spolaor10,kuntschner10,tortora10,lb10} have found metallicity gradients 
for elliptical galaxies have a minimum around $M_*=10^{10.3}M_{\odot}$. We also seem to see a minimum 
median metallicity gradient around this stellar mass in Fig. \ref{gradvsm} (row 3, the rightmost column).
However, this  trend in Fig. \ref{gradvsm} is a combination of  spirals and ellipticals
with gradients derived within 0.5-1.5$R_e$.
In Fig. \ref{comparecalifa}, neither results from CALIFA \citep{gd15} nor our results show this trend.

\begin{figure} 
\begin{center}
\includegraphics[scale=0.4]{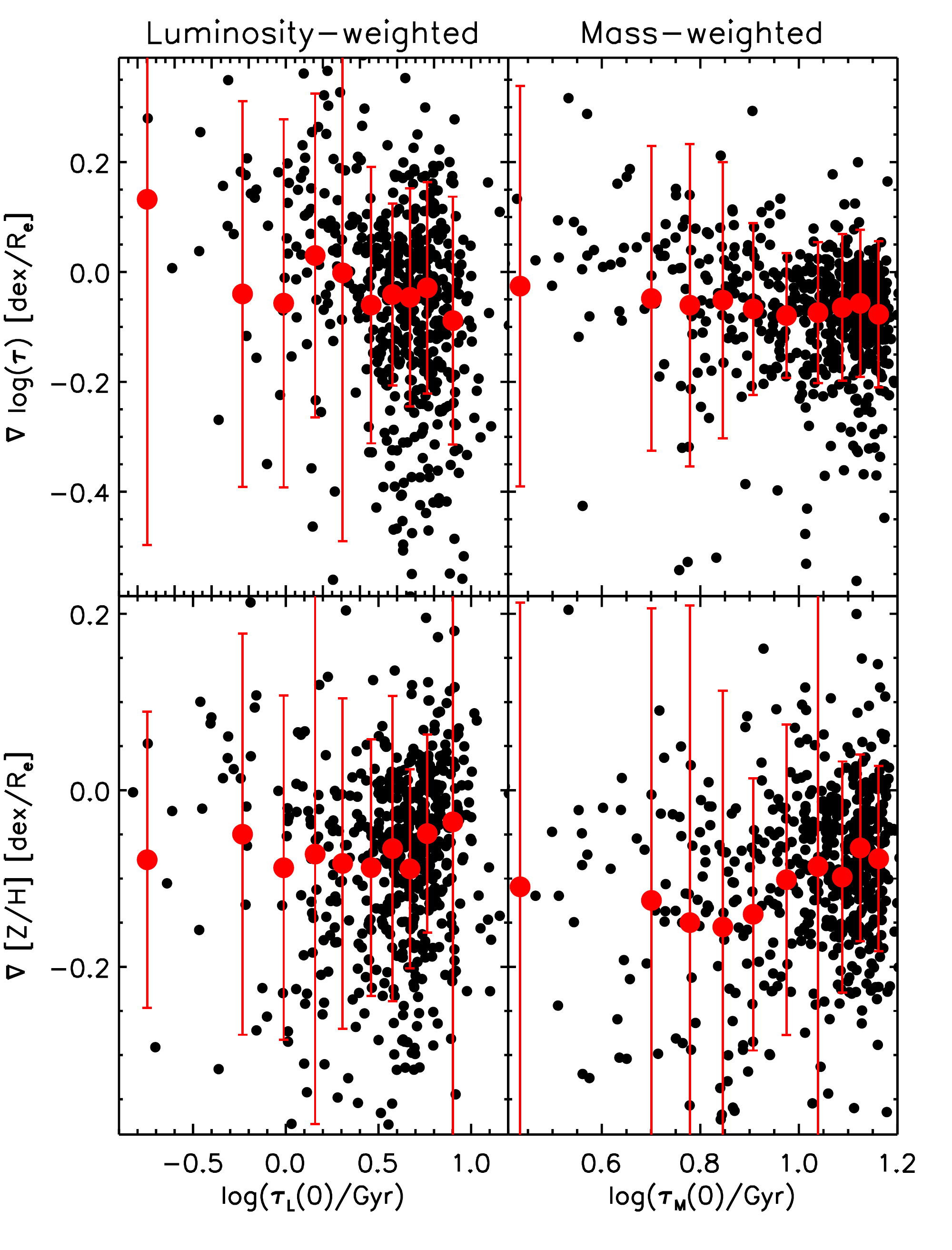}  

\caption{Central ages versus gradients for elliptical galaxies. Left panel show the luminosity-weighted parameters and right panels show the mass-weighted parameters. Black dots show individual galaxies and the big red dots with error bars show median values with $1\sigma$ scatter of each central age bin.}
\label{zpcenvsgrads}
\end{center}
\end{figure}

Also, many studies found that younger elliptical galaxies have more negative metallicity 
gradients and more positive age gradients than older ones
\citep[e.g.][]{hopkins09,sb09,tortora10, rawle10}. 
Here we plot the central age versus ages and metallicity gradients for elliptical galaxies from our sample in 
Fig. \ref{zpcenvsgrads}. The lower-right panel of Fig. \ref{zpcenvsgrads} shows a weak increasing trend with central age, which is qualitatively consistent with previous results. However, the scatters are large. Other panels do not show this trend.

\citet{lb05} studied colour gradients of  1700 early-type galaxies in 159 galaxy clusters and found that colour gradients 
strongly depend on cluster richness. 
More recently, \citet{tamura00} and \citet{to00, to03} studied colour gradients of early-type galaxies in clusters and less dense environments and found that the colour gradients do not depend on environment.  
\citet{tortora11} studied stellar population gradients for a sample of group and cluster galaxies from numerical simulations 
and found that for low-mass galaxies, age gradients of cluster galaxies are higher (more positive) 
than those of group galaxies whilst metallicity gradients of cluster galaxies are lower (more negative)
than those of group galaxies. The situation is inverted for high-mass galaxies.  
However, these trends look weak (mostly within 1$\sigma$ uncertainty), especially for age gradients. 
Our results show  weak or nearly no dependence on both LSS and local density environments and
are  consistent with \citet{tamura00} and \citet{to00, to03}.

As mentioned in previous sections, we have a one companion MaNGA paper (Goddard et al. 2016a) studying stellar age and metallicity gradients versus galaxy local density environment measured by close neighbours, and another companion MaNGA paper studying stellar population gradients versus galaxy properties (Goddard et al. 2016b). The two companion papers use different methods from this paper: they use different software \citep[FIREFLY,][]{wilkinson15} and different stellar population models \citep{ms11} to derive stellar age and metallicity gradients.  These differences result in slightly more positive mass-weighted age gradients than ours (see details in Goddard et al. 2016b). Also, they only use the MaNGA Primary sample, which has 721 galaxies and focus on 0-1.5 Re region of each galaxy. Despite these differences, their results are similar to ours, i.e. stellar age and metallicity gradients do not have an obvious correlation with local density environment (Goddard et al. 2016a).

\subsubsection{Values at $R_e$}
There have been many studies about the environmental dependence of overall galactic properties. 
Although overall galactic properties are different from values at $R_e$, they should be correlated given that
age and metallicity gradients are shallow for most galaxies. 
For example, \citet{ll08} studied overall galactic properties in 
different LSS environments and found that at fixed luminosity ($9.4 < \log(M_*/M_{\odot}) < 10$), 
galaxies in sheet and void environments have lower
D(4000) values than those in cluster environment. 
This implies that galaxies in void and sheet 
environments have younger luminosity-weighted ages and is consistent with our results shown
in Fig. \ref{zeropvsenv}. The difference is that we further found that galaxies with smaller stellar masses
are more affected by LSS environments.
\citet{kauffmann04} studied correlations between galaxy properties and local density environments 
and found that $D_n(4000)$ depends strongly on local density and that the dependence is strongest for
low mass galaxies. This is also consistent with our findings shown in Fig. \ref{zeropvslocalden}.
Mass-weighted ages are more affected by low mass stars and  therefore show smaller differences. 
Possible mechanisms causing the differences could be harassment \citep[e.g.][]{moore98}, strangulation \citep[e.g.][]{balogh00, peng15}, 
or gas stripping \citep[e.g.][]{gg72}, which could suppress recent star formation in low mass galaxies in dense environments.
We defer further investigation of environmental affects to a future paper, which will use a larger galaxy sample 
and include halo merger history information from \citet{wang16}.

\subsection{Implications}
Fig. \ref{gradvsenv} shows that age and metallicity gradients have weak or no dependence on LSS. There may be some environmental dependence for low-mass and blue galaxies (e.g. the last row of Fig. \ref{gradvsenv}). The differences in different LSS environments however, are within a $2\sigma$ error. A larger galaxy sample is needed to confirm or disprove this  environmental dependence.  For high-mass and red galaxies, there are no obvious dependences on LSS environments. 

This implies that the galaxy stellar population structure is more affected by previous mergers or by internal processes such as radial migrations. 
The former could be checked by examining the correlation between stellar population gradients and galaxy merger histories. We are currently determining the merger histories for the dark haloes of all MaNGA galaxies following \citet{wang14}, as metioned in Section \ref{envmethod}, and  will present the results in a future paper. 
The importance of  internal process is becoming more and more recognized. In particular, more recent studies \citep{roskar08, sb09, minchev12, zheng15} show that  stellar radial migration plays an important role in shaping  disk galaxies. 
The co-added metallicity profiles (Fig. \ref{mzl_prof} and \ref{mzm_prof}) show an upturn in the region $>1.5\,  R_e$, while the luminosity-weighted age profiles of disk galaxies (upper row of Fig. \ref{agel_prof}) show a `U' shape with the minimum located around $1-1.5\,  R_e$.  This is  similar to the results found by \citet{zheng15} using multi-band photometric data of 700 disk galaxies \citep[see also][etc.]{bakos08, azzollini08, rs12, yoachim12,  herrmann13, marino16, rl16}, and may be a signature of stellar radial migration as seen in numerical simulations by \citet{roskar08}. 
The `U' shaped age profile is not found in elliptical galaxies \citep[see Figs. 4-5 and][]{gd15}. This might be because ellipticals do not have structures like bars or spiral arms, which are crucial for stellar radial migration models proposed in the literature \citep{sb02}. 
However, there could also be other explanations \citep[e.g.][]{sb09, rl16}. We have initiated a further project to study the `U' shaped age profile using  MaNGA data.

\begin{table*}

\caption{Comparison with other studies. Note -- MW in the Comments column means mass-weighted and $r_{d0}$ in the Radial range column means the radius at which the light starts being dominated by the disk \citep{sb14}.  The last column shows the references: R10 - \citet{rawle10}; K10 - \citet{kuntschner10}; SB14 - \citet{sb14}; GD15 - \citet{gd15}; G - our companion paper (Goddard et al. 2016b, submitted). This table is intended to compare our results with a few recent studies, not a complete list of all previous studies. We only use the MaNGA primary and secondary sample galaxies in calculating the mean value and uncertainties listed in this table because other galaxies do not have a well defined volume weight. We apply volume weight correction according to the MaNGA sample selection paper Wake et al. (in prep.).
}
\label{compare}
\begin{center}
\begin{tabular}{cccccccc}
\hline
\hline

Galaxy type &  Number &  $\triangledown \log(\tau$) & $\triangledown$ [Z/H] & Units &  Radial range & Comments & Ref.  \\
\hline
Disk &         422 & $-0.08\pm0.02$ & $-0.14\pm0.02$ &  \multirow{2}{*}{dex/$R_e$} & \multirow{2}{*}{0.5-$1.5R_e$}  & \multirow{2}{*}{MW}  & \multirow{2}{*}{This paper} \\
Elliptical &         463 & $-0.05\pm0.01$ & $-0.09\pm0.01$ &  &  & \\
 \hline
Disk &         422 & $-0.18\pm0.03$ & $-0.31\pm0.04$ & \multirow{2}{*}{dex/dex} & \multirow{2}{*}{0.5-$1.5 R_e$} & \multirow{2}{*}{MW}  & \multirow{2}{*}{This paper} \\
Elliptical &         463 & $-0.13\pm0.02$ & $-0.19\pm0.03$ \\
\hline
Early type & 25 & $-0.02\pm0.06$ & $-0.13\pm0.04$ & dex/dex & 0-$1.5 R_e$ & cluster & R10\\
\hline
\multirow{2}{*}{Early type}  & \multirow{2}{*}{48}  & $0.02\pm$0.13 & $-0.25\pm0.11$ & \multirow{2}{*}{dex/dex}  & \multirow{2}{*}{$2''$ - $1R_e$} & old & \multirow{2}{*}{K10} \\
 &  & $0.28\pm0.16$ & $-0.28\pm0.12$ & &  & young &  \\
 \hline
Disk 	& 62 & $0.000\pm0.006$ & $-0.087\pm0.008$ & dex/$R_e$ & $r_{d0}$ - $1.5R_e$ &  MW & SB14 \\
\hline
All type & 300 & $-0.4$ - $0.4$ & $-0.1\pm0.15$ & dex/$R_e$ &  0-1$R_e$ &  MW & GD15 \\
 \hline
 Disk & 216 & $0.07 \pm 0.07$ & $-0.102 \pm 0.1$ & \multirow{2}{*}{dex/$R_e$} & \multirow{2}{*}{0-$1.5 R_e$} & \multirow{2}{*}{MW} & \multirow{2}{*}{G} \\
 Elliptical & 505 &  $0.09\pm0.05$ & $-0.05\pm0.07$   \\
\hline
\hline

\end{tabular}
\end{center}
\end{table*}%

\section{Conclusions}
\label{conclusions}
We have studied the stellar age and metallicity distributions of 1005 galaxies from the MaNGA MPL-4 internal data release \citep[equivalent to SDSS DR13 public release,][]{sdssdr13}. 
We have derived the age and metallicities by applying the STARLIGHT package to MaNGA IFU spectra. 
We have obtained the age and metallicity gradients of each galaxy by fitting a straight line to their age and metallicity profiles over 0.5-1.5$R_e$,  and have explored their correlations with  total stellar mass $M_*$, $NUV-r$ colour and two different environment indicators using the large scale tidal field and the local density. 

We found the mean age and metallicity gradients are close to zero but slightly negative: mean metallicity gradient $\triangledown [Z/H]_M=  -0.14\pm0.02$\, dex/$R_e$ for disk galaxies and $\triangledown [Z/H]_M= -0.09\pm0.01$\, dex/$R_e$ for elliptical galaxies; mean age gradient $\triangledown \tau_M=  -0.08\pm0.02$\, dex/$R_e$ for disk galaxies and $\triangledown \tau_M=  -0.05\pm0.01$\, dex/$R_e$ for elliptical galaxies. This is consistent with the inside-out formation scenario.  The zero but slightly negative gradient is seen as an average over many galaxies however,  gradients for individual galaxies can be positive, negative or zero.

We found that both the age and metallicity gradients have weak or no dependence on either the large scale structure (LSS) or the local density environment in the context of our current galaxy sample.  

As a complementary investigation, we have also studied correlations between age and metallicity values at the effective radii, and galaxy overall properties as well as environments. Age and metallicity  are highly correlated with  stellar mass, $NUV-r$ colour and LSS environment and the local density.  Low-mass galaxies tend to be younger and have lower metallicity in low-density environments while high-mass galaxies are less affected by environment. 

In conclusion,  internal processes in galaxy evolution history appear to play a major role in shaping  galaxies, especially the high-mass and red galaxies.

\section*{Acknowledgements}
We thank Claudia Maraston and Brice M$\rm \acute{e}$nard for helpful discussions and suggestions. We also thank the referee for very helpful suggestions.

This work is supported by the Young Researcher Grant of National Astronomical Observatories, Chinese Academy of Sciences (ZZ), and  the Strategic Priority Research Programme ``The Emergence of Cosmological Structures" of the Chinese Academy of Sciences, Grant No. XDB09000000 (SM, CL, and RJL). SM is also supported by NSFC (grant No. 11333003, 11390372). 
CL is also supported by National Key Basic Research Program of China (No. 2015CB857004), NSFC (grant No. 11173045, 11233005, 11325314, 11320101002).
HW is supported by NSFC (11522324,11421303).
KB is supported by World Premier International Research Center Initiative (WPI Initiative), MEXT, Japan and by  JSPS KAKENHI Grant Number 15K17603. 
DG acknowledges support from an STFC studentship.
DB acknowledges support from grant RSF 14-50-00043.
RR thanks to CNPq and FAPERGS for financial support.

This work makes use of data from SDSS-IV.
Funding for SDSS-IV has been provided by the Alfred P. Sloan Foundation and Participating Institutions. 
Additional funding towards SDSS-IV has been provided by the U.S. Department of Energy Office of Science. SDSS-IV
acknowledges support and resources from the Centre for High-Performance Computing at the University of Utah. The SDSS web site is www.sdss.org.

SDSS-IV is managed by the Astrophysical Research Consortium for the Participating Institutions of the SDSS Collaboration including the Brazilian Participation Group, the Carnegie Institution for Science, Carnegie Mellon University, the Chilean Participation Group, the French Participation Group, Harvard-Smithsonian Center for Astrophysics, Instituto de Astrof'sica de Canarias, The Johns Hopkins University, Kavli Institute for the Physics and Mathematics of the Universe (IPMU) / University of Tokyo, Lawrence Berkeley National Laboratory, Leibniz Institut f$\ddot{\rm u}$r Astrophysik Potsdam (AIP), Max-Planck-Institut f$\ddot{\rm u}$r Astronomie (MPIA Heidelberg), Max-Planck-Institut f$\ddot{\rm u}$r Astrophysik (MPA Garching), Max-Planck-Institut f$\ddot{\rm u}$r Extraterrestrische Physik (MPE), National Astronomical Observatory of China, New Mexico State University, New York University, University of Notre Dame, Observat$\acute{\rm a}$rio Nacional / MCTI, The Ohio State University, Pennsylvania State University, Shanghai Astronomical Observatory, United Kingdom Participation Group, Universidad Nacional Aut$\acute{\rm o}$noma de M$\acute{\rm e}$xico, University of Arizona, University of Colorado Boulder, University of Oxford, University of Portsmouth, University of Utah, University of Virginia, University of Washington, University of Wisconsin, Vanderbilt University,
and Yale University.








\appendix

\section{Morphology and central/satellite effect}

Different morphologies (disk versus elliptical) and different environments within a galaxy group 
(e.g. central versus satellite) may also affect the age and colour gradients.
To test this, we have split our sample into subsamples of disk-like (S{\'e}rsic index $n<2.5$) and elliptical-like ($n\ge 2.5$)
galaxies, as well as central and satellite galaxies, and produce plots similar to 
Figs. \ref{gradvsenv} and \ref{zeropvsenv}. Note that some of the results shown here are not robust 
because the number of galaxies in each bin is small, typically much less than 30.

\begin{figure*} 
\begin{center}
\includegraphics[scale=0.37]{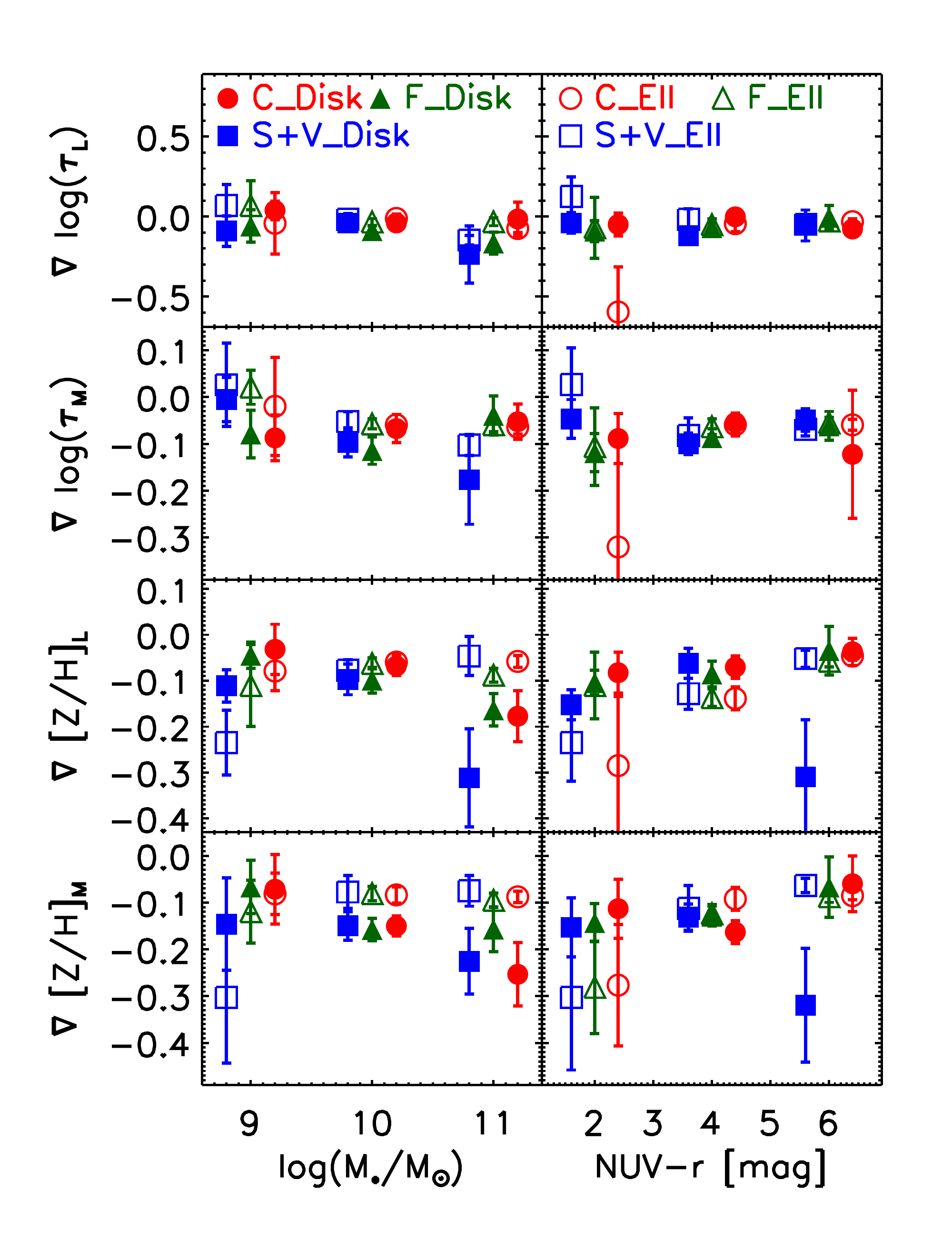}  
\includegraphics[scale=0.37]{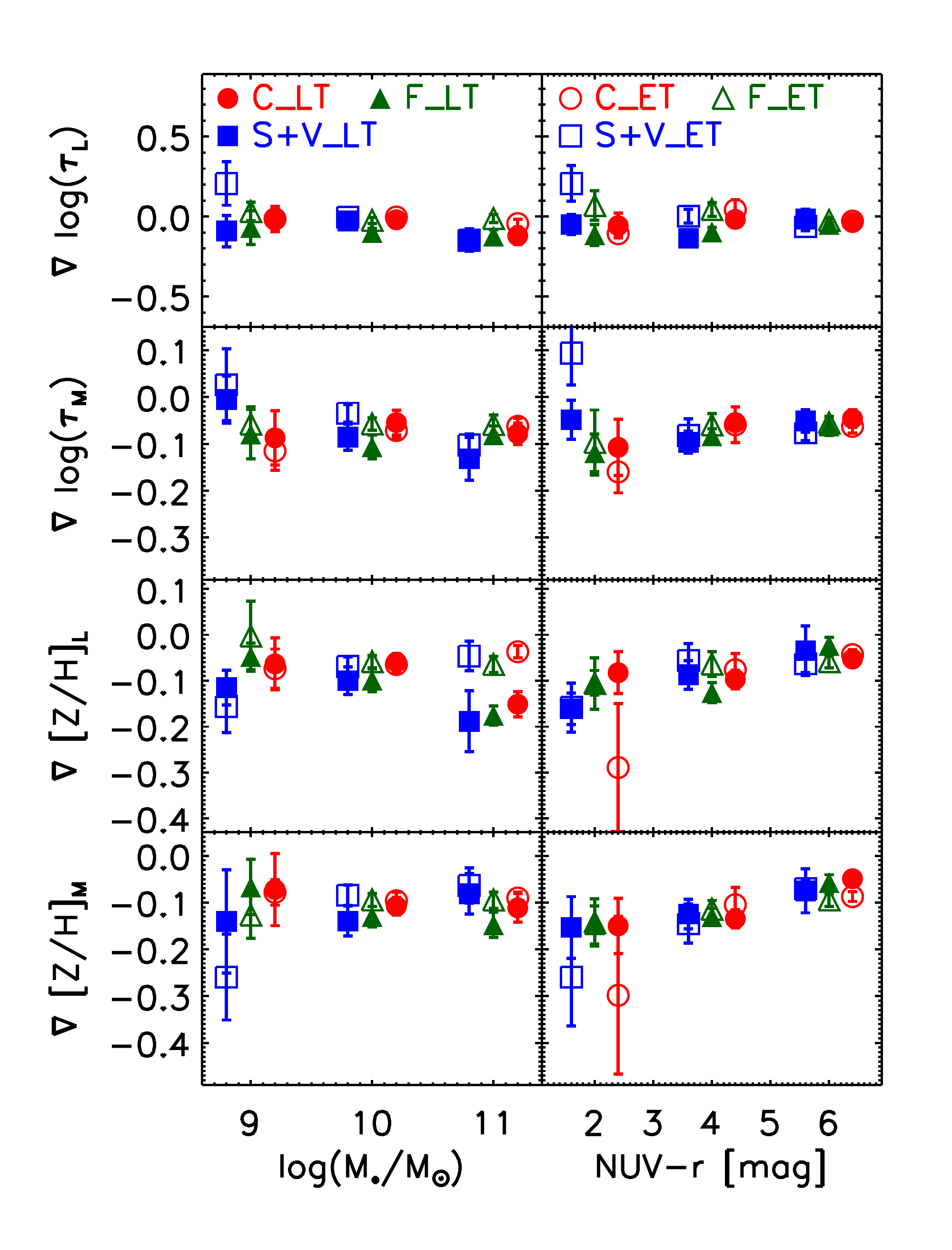}  
\caption{Age and metallicity gradients in each mass and $NUV-r$ colour bin. Red dots show cluster environment (C),  green triangles show filament environment (F), and blue squares show sheet and void environment (S+V). Filled symbols show late type/disk galaxies (LT/Disk) and open symbols show early type/elliptical galaxies  (ET/Ell). Left panels show results using morphological classification based on S{\`e}rsic index $n$ and right panels show results using morphological classification based on visual classification. 
}
\label{gradvsenvmorph}
\end{center}
\end{figure*}

\begin{figure*} 
\begin{center}
\includegraphics[scale=0.37]{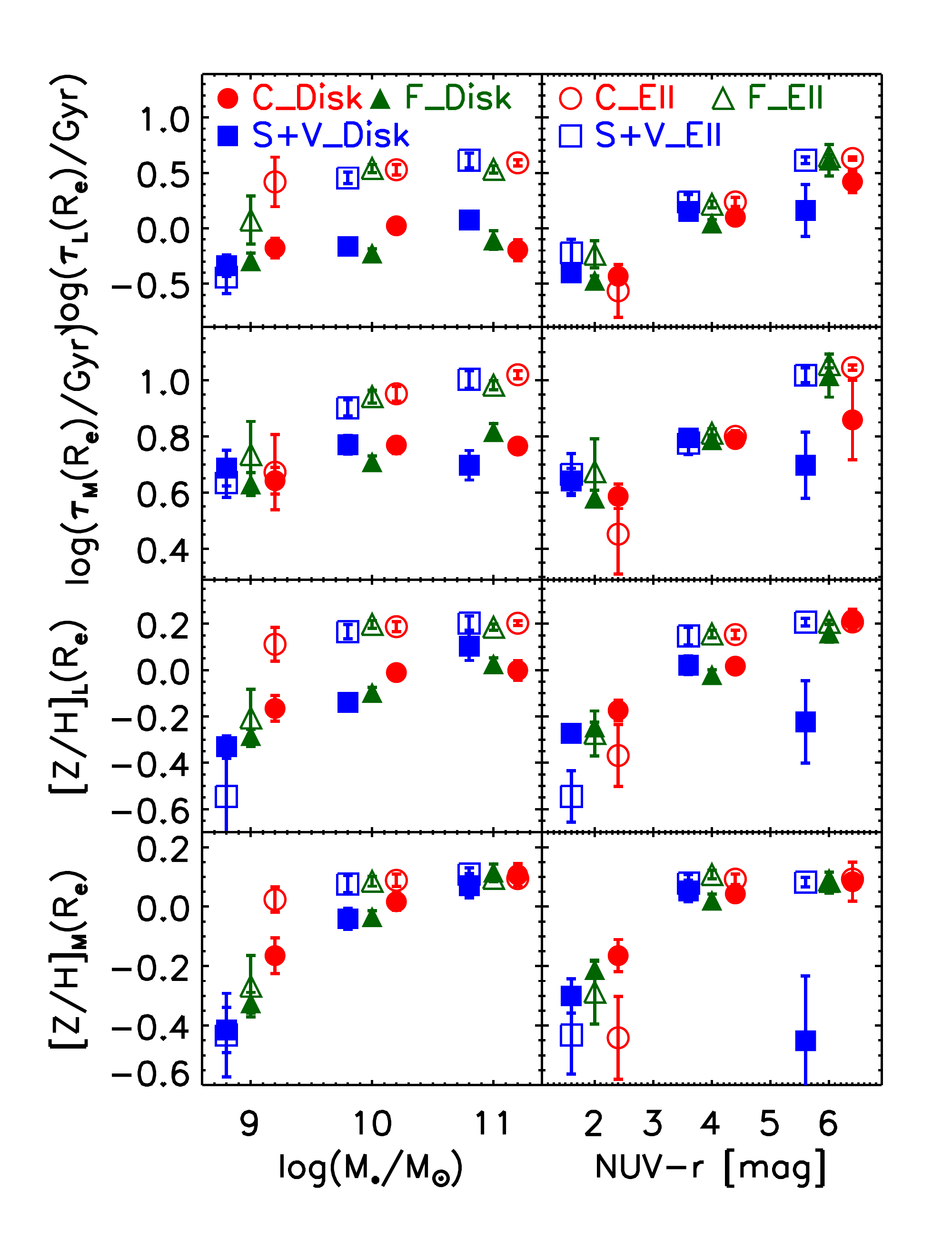}  
\includegraphics[scale=0.37]{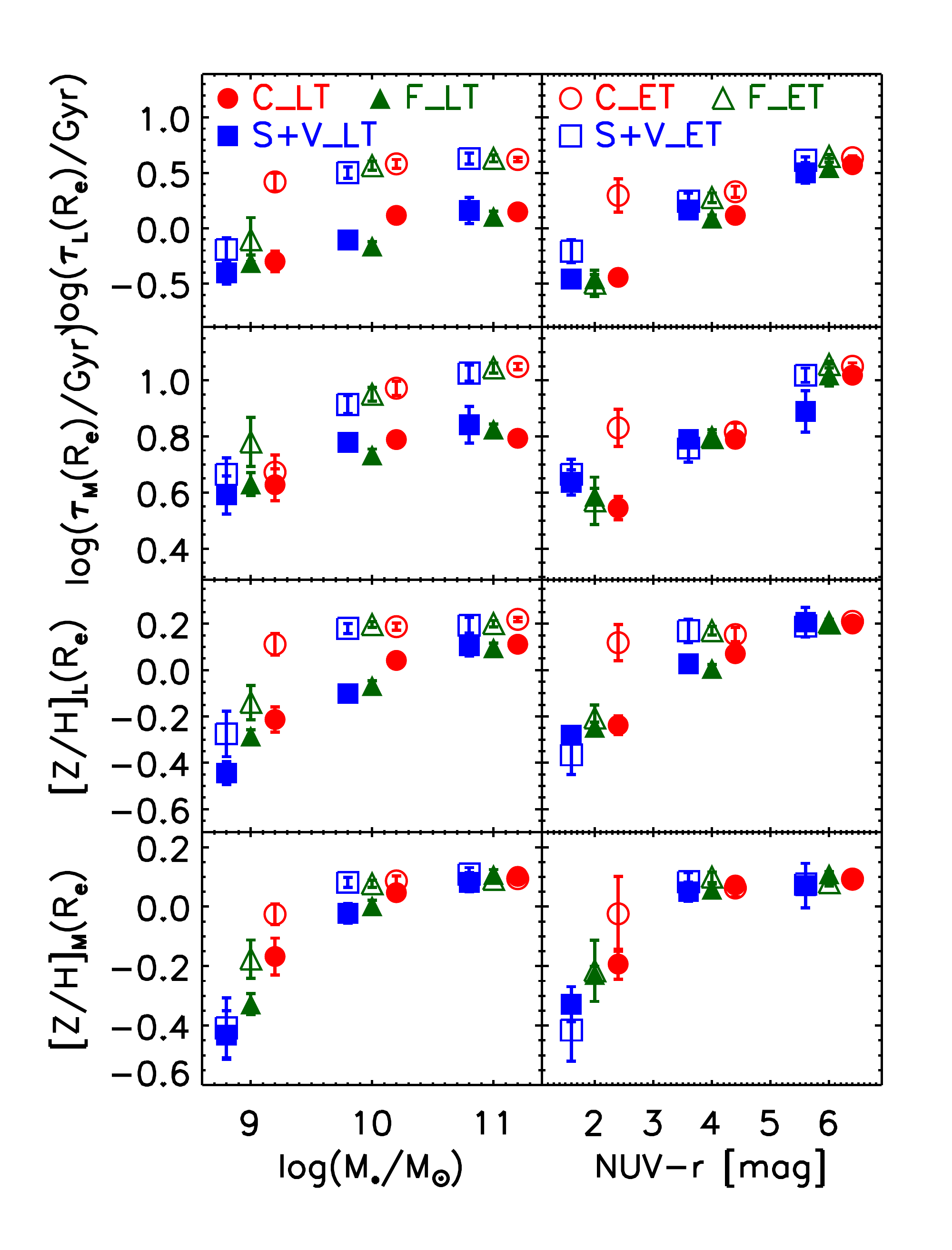} 
\caption{Median values of fitted age and metallicity at the effective radius $R_e$ in each mass and $NUV-r$ colour bin. Panel arrangement and symbols are the same as Fig. \ref{gradvsenvmorph}. 
}
\label{zeropvsenvmorph}
\end{center}
\end{figure*}

We present the morphology effect in the left panels of Figs. \ref{gradvsenvmorph} and \ref{zeropvsenvmorph}. 
It can be seen that ellipticals usually have weak gradients  (closer to zero). 
They are also older and more metal rich than disk galaxies. However, the differences between 
different LSS environments are small and are mostly within the $1\sigma$ uncertainty. 

Furthermore, we also performed the morphological classification using visual inspection. 
We visually inspected about 1000 galaxies from the MaNGA MPL-4 sample and classify them
as late type (Sa, Sb, Sc, Sd, and Im) or early type (E and S0) based on the method described in \citet{na10}.
The results using this classification scheme are presented in the right panels of Figs. \ref{gradvsenvmorph} and \ref{zeropvsenvmorph}.
For our purpose, it is clear that the two morphological classification schemes (using S{\'e}rsic index or visual inspection)
lead to very similar results (except for some bins with really small numbers of galaxies).

\citet{sb06b} studied the central regions of 98 early type galaxies and found that 
galaxies in low-density environments appear younger and more metal rich than their counterparts in 
high-density environments. This appears somewhat different from our results here: we find low-mass 
elliptical galaxies in sheet and void regions are both younger and metal poorer than their counterparts 
in clusters and filaments, and we see no difference for high-mass galaxies between different 
LSS environments. (Fig. \ref{zeropvsenvmorph}). It should be noted, however, that the
environment and metallicity estimators used here are different from those in \citet{sb06b}. 
In addition, while they focussed on the central parts of galaxies, our results  are 
for regions near $R_e$.

\begin{figure} 
\begin{center}
\includegraphics[scale=0.37]{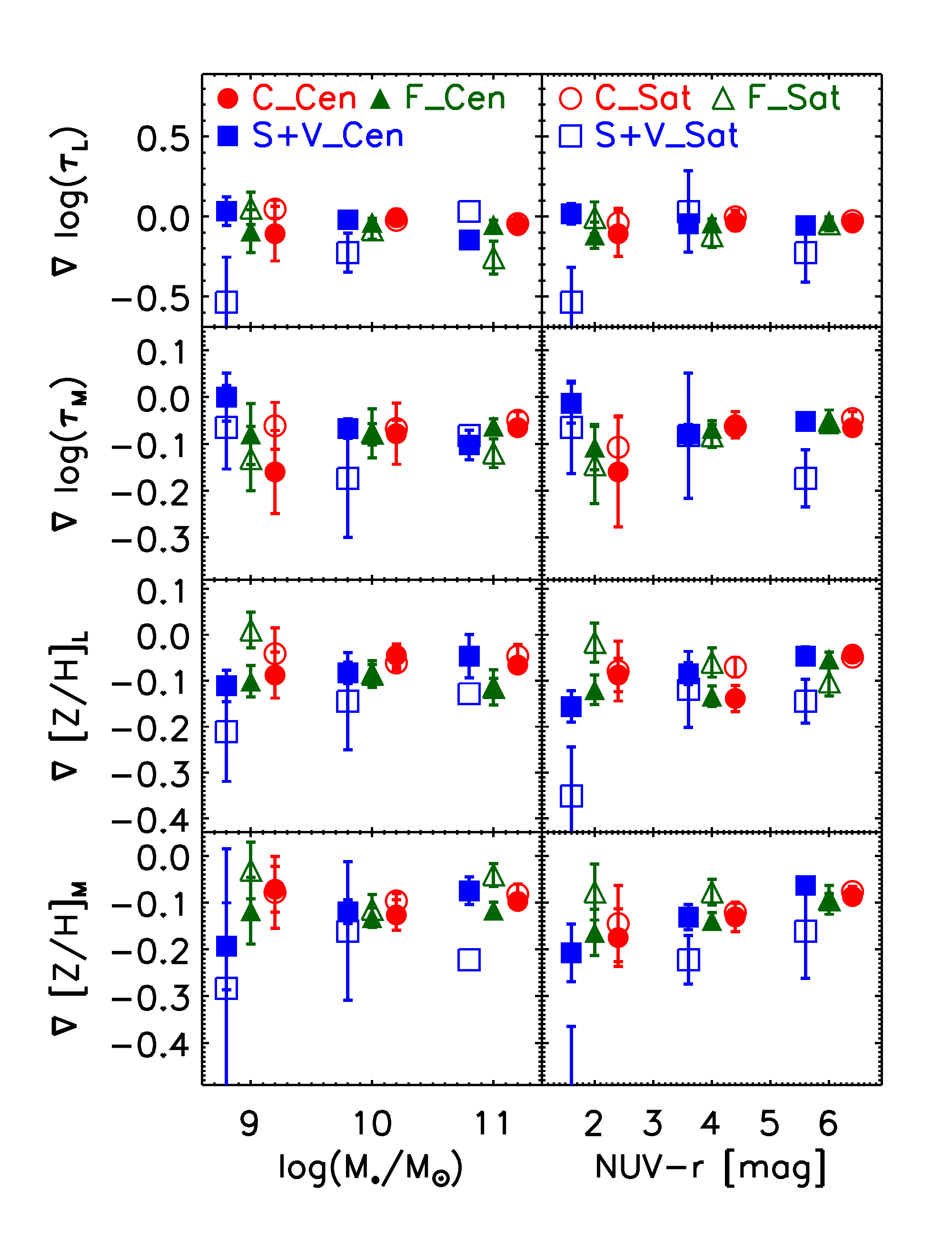}  
\vspace*{-0.4cm }
\caption{Age and metallicity gradients in each mass (left panels) and $NUV-r$ colour (right panels) bin. Red dots show cluster environment (C),  green triangles show filament environment (F), and blue squares show sheet and void environment (S+V). Filled symbols show central galaxies (Cen) and open symbols show satellite galaxies (Sat).
}
\label{gradvsenvcs}
\end{center}
\end{figure}

\begin{figure} 
\begin{center}
\includegraphics[scale=0.37]{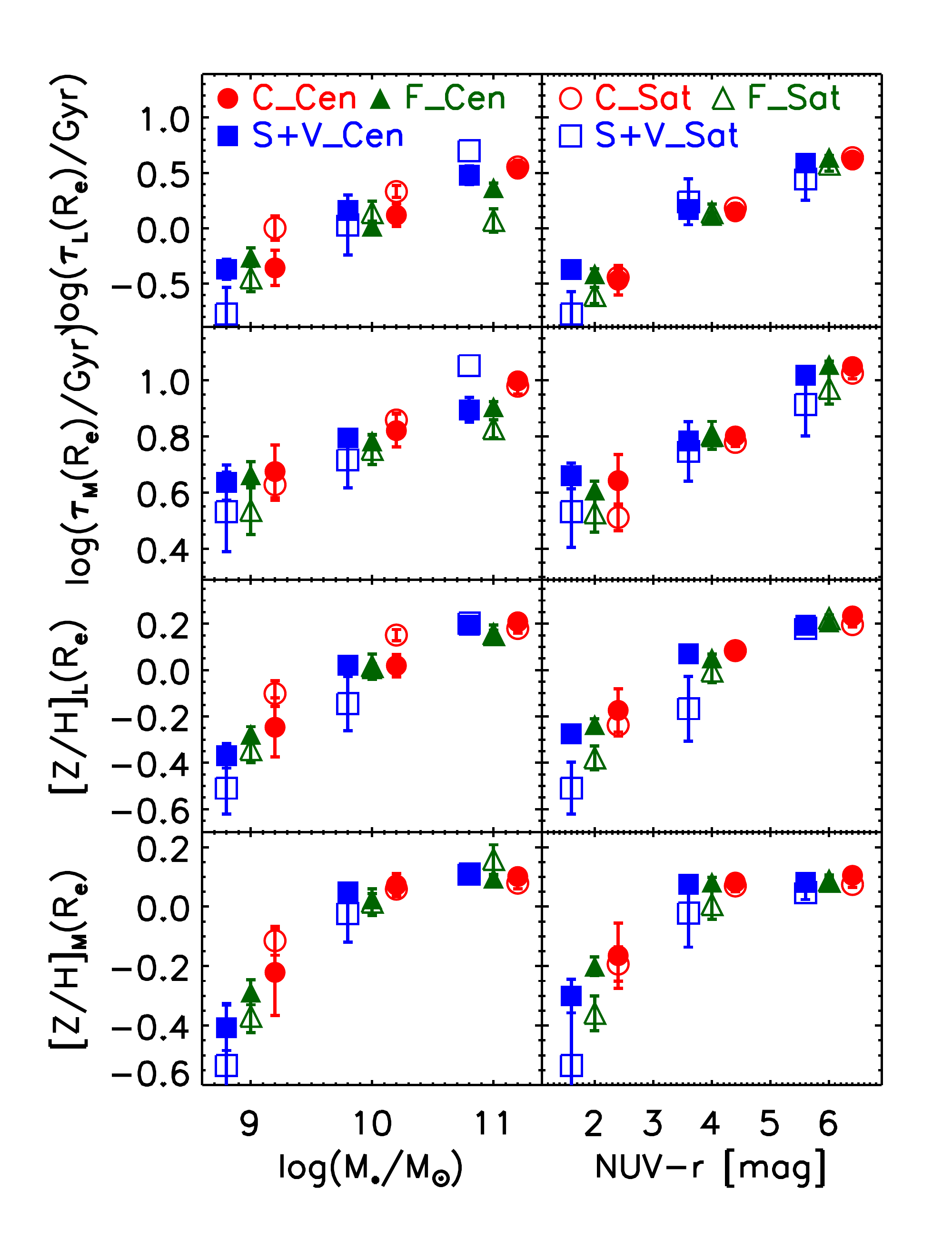}  
\vspace*{-0.4cm }
\caption{Median values of fitted age and metallicity at the effective radius $R_e$ in each mass (left panels) and $NUV-r$ colour (right panels) bin. Symbols are the same as Fig. \ref{gradvsenvcs}. 
}
\label{zeropvsenvcs}
\end{center}
\end{figure}

We also use the central (the most massive galaxy within a galaxy group) and satellite information from the catalogue of \citet{yang07} to study the central v.s. satellite effect. 
The results are plotted in Fig. \ref{gradvsenvcs} and \ref{zeropvsenvcs}. Here again, no significant difference is seen between gradients of centrals and satellites.  
\citet{vdb08} studied colour and morphology differences between centrals and  satellites and found that satellites are redder than centrals of the same stellar mass. 
\citet{fg15}  found early type satellites are slightly (0.02 dex) older  than early type centrals. \citet{pasquali10} conducted a more detailed study and
found that satellites are older and metal richer than centrals of the same stellar mass, and that this difference increases with decreasing stellar mass. 
They also found that the differences are less in denser galaxy environments. 
We observed a similar trend for galaxies in cluster environments (Fig. \ref{zeropvsenvcs}).
 We found satellite galaxies in sheet and void environments are relatively younger and 
metal poorer compared to centrals in the same LSS environments. 
However, these differences shown in our plots are small by comparison with the error bars.

\section{Beam smearing effect}

\begin{figure} 
\begin{center}
\includegraphics[scale=0.33]{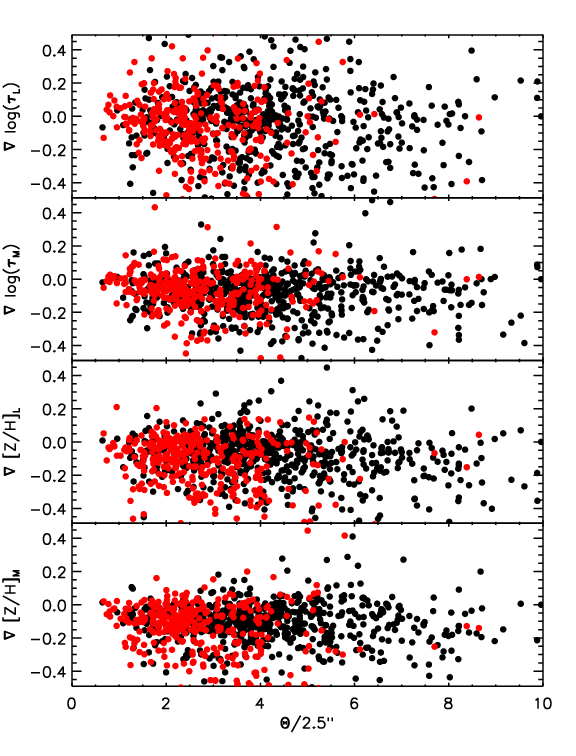}  
\vspace*{-0.3cm }
\caption{ Age and metallicity gradients versus galaxy angular size $\Theta$ ($=1.5\,R_e$) in units of beam size ($2.5''$). 
Black dots are for galaxies from the MaNGA primary+color-enhanced sample and red dots are for galaxies from the secondary sample.
This diagram shows that the gradients do not depend on the galaxy angular size.}
\label{gradvssmear}
\end{center}
\end{figure}

\begin{figure} 
\begin{center}
\includegraphics[scale=0.33]{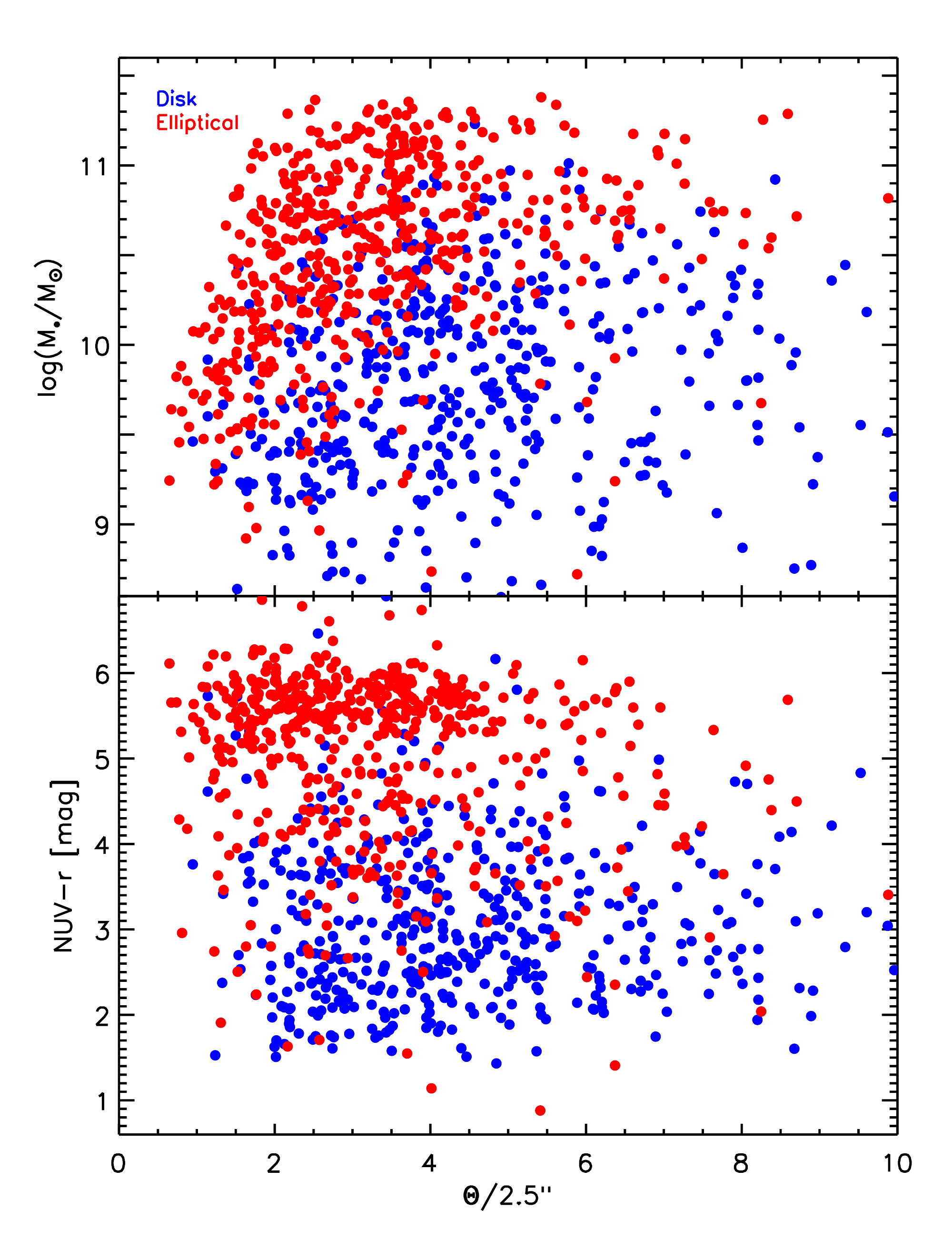}  
\vspace*{-0.3cm }
\caption{ Galaxy overall properties $M_*$ and NUV$-r$ colour versus galaxy angular size $\Theta$ in units of beam size ($2.5''$). Blue dots are for galaxies with S$\rm \acute{e}$rsic indices $n< 2.5$ (disk galaxies),
and red dots are for galaxies with S$\rm \acute{e}$rsic indices $n\ge 2.5$ (elliptical galaxies).}
\label{galpropvssmear}
\end{center}
\end{figure}

One concern about deriving age and metallicity gradients from MaNGA data is that
the gradients might be affected by beam smearing effects. One way to examine the effect of beam
smearing is to plot our derived gradients versus angular sizes of the gradient-fitting region (Fig. \ref{gradvssmear}). 
Ideally we would expect no correlation between gradients and angular sizes if any beam smearing 
effect is small. We can see from Fig. \ref{gradvssmear} that there is no obvious correlation 
between gradients and galaxy angular sizes. The weak trend of luminosity-weighted age gradients of
MaNGA secondary sample galaxies is dominated by stellar mass. As we can see from 
Fig. \ref{galpropvssmear}  high-mass galaxies tend to have larger angular sizes. 
Therefore, beam smearing effects should not be a big problem for our purposes.

\bsp	
\label{lastpage}
\end{document}